%% file: main.tex
\DocumentMetadata{}
\documentclass[sigconf]{acmart}

\setcopyright{rightsretained}
\copyrightyear{2024}
\acmYear{2024}
\acmConference[ASE '24]{39th IEEE/ACM International Conference on Automated
Software Engineering }{October 27-November 1, 2024}{Sacramento, CA, USA}
\acmBooktitle{39th IEEE/ACM International Conference on Automated Software
Engineering (ASE '24), October 27-November 1, 2024, Sacramento, CA, USA}
\acmDOI{10.1145/3691620.3695530}
\acmISBN{979-8-4007-1248-7/24/10}

\input{feng_math_cmds}

\newcommand{\todoc}[2]{{\textcolor{#1}{\textbf{#2}}}}
\newcommand{\todored}[1]{{\todoc{red}{\textbf{[#1]}}}}
\newcommand{\todogreen}[1]{\todoc{green}{\textbf{[#1]}}}

\newcommand{\todoorange}[1]{\todoc{orange}{\textbf{[#1]}}}
\newcommand{\todobrown}[1]{\todoc{brown}{\textbf{[#1]}}}

\newcommand{\todopurple}[1]{\todoc{purple}{\textbf{[#1]}}}
\newcommand{\todocyan}[1]{\todoc{cyan}{\textbf{[#1]}}}
\newcommand{\todoteal}[1]{\todoc{teal}{\textbf{[#1]}}}

\newif\ifclear
\clearfalse
\ifclear
\newcommand{\todo}[1]{}
\newcommand{\xz}[1]{}
\newcommand{\sw}[1]{}
\newcommand{\xx}[1]{}
\newcommand{\hc}[1]{}
\newcommand{\yy}[1]{}
\newcommand{\zc}[1]{}
\newcommand{\qs}[1]{}

\else
\newcommand{\todo}[1]{\todocyan{TODO: #1}}
\newcommand{\xz}[1]{\todored{XZ: #1}}
\newcommand{\sw}[1]{\todoorange{SF: #1}}
\newcommand{\xx}[1]{\todopurple{XX: #1}}
\newcommand{\hc}[1]{\todoteal{HC: #1}}
\newcommand{\yy}[1]{\todocyan{YY: #1}}
\newcommand{\zc}[1]{\todogreen{ZC: #1}}
\newcommand{\qs}[1]{\todobrown{QS: #1}}

\fi

\newcommand{\toolname}{{\sc Rocas}\xspace}
\newcommand{\var}[1]{\texttt{#1}}
\newcommand{\func}[1]{{\texttt{#1}}}
\newcommand{\metric}[1]{{$\mathtt{#1}$}}

\AtBeginDocument{%
  \providecommand\BibTeX{{%
    \normalfont B\kern-0.5em{\scshape i\kern-0.25em b}\kern-0.8em\TeX}}}

\begin{document}

\title[ROCAS: Root Cause Analysis of Autonomous Driving Accidents via Cyber-Physical Co-mutation]{ROCAS: Root Cause Analysis of Autonomous Driving Accidents via Cyber-Physical Co-mutation}

\author{Shiwei Feng}
\affiliation{
  \institution{Purdue University}
  \city{West Lafayette}
  \country{USA}
}
\email{feng292@purdue.edu}
\orcid{0000-0001-6959-4327}

\author{Yapeng Ye}
\affiliation{
  \institution{Purdue University}
  \city{West Lafayette}
  \country{USA}
}
\email{ye203@purdue.edu}
\orcid{0000-0001-7232-0650}

\author{Qingkai Shi}
\authornote{Corresponding author.}
\affiliation{
  \institution{\hspace{-2mm}The State Key Laboratory for Novel Software Technology, Nanjing University}
  \city{Nanjing}
  \country{China}
}
\email{qingkaishi@nju.edu.cn}
\orcid{0000-0002-8297-8998}

\author{Zhiyuan Cheng}
\affiliation{
  \institution{Purdue University}
  \city{West Lafayette}
  \country{USA}
}
\email{cheng443@purdue.edu}
\orcid{0000-0001-7280-6079}

\author{Xiangzhe Xu}
\affiliation{
  \institution{Purdue University}
  \city{West Lafayette}
  \country{USA}
}
\email{xu1415@purdue.edu}
\orcid{0000-0001-6619-781X}

\author{Siyuan Cheng}
\affiliation{
  \institution{Purdue University}
  \city{West Lafayette}
  \country{USA}
}
\email{cheng535@purdue.edu}
\orcid{0009-0006-0903-6917}

\author{Hongjun Choi}
\affiliation{
  \institution{DGIST}
  \city{Daegu}
  \country{South Korea}
}
\email{hongjun@dgist.ac.kr}
\orcid{0000-0003-4706-934X}

\author{Xiangyu Zhang}
\affiliation{
  \institution{Purdue University}
  \city{West Lafayette}
  \country{USA}
}
\email{xyzhang@cs.purdue.edu}
\orcid{0000-0002-9544-2500}

\renewcommand{\shortauthors}{S. Feng, Y. Ye, Q. Shi, Z. Cheng, X. Xu, S. Cheng, H. Choi and X. Zhang}

\begin{abstract}
\input{0-abstract}
\end{abstract}

\maketitle

\input{1-introduction}

\input{2-background}

\input{3-motivation}

\input{4-definition}

\input{5-design}

\input{6-evaluation}

\input{7-discussion}

\input{8-related}

\input{9-conclusion}

\newpage
\bibliographystyle{ACM-Reference-Format}
\bibliography{reference.bib}

\input{10-appendix}

\end{document}

%% file: feng_math_cmds.tex
\usepackage[utf8]{inputenc} %
\usepackage[T1]{fontenc}    %

\usepackage{url}            %

\usepackage{breakurl}
\usepackage{hyperref}       %
\usepackage{nicefrac}       %
\usepackage{microtype}      %
\usepackage{xcolor}         %
\usepackage{graphicx, wrapfig, blindtext}
\usepackage{multirow}
\usepackage{multicol}
\usepackage{makecell}
\usepackage{subcaption}
\usepackage{bibunits}

\usepackage{booktabs}   %

\usepackage{latexsym}
\usepackage{xspace}
\usepackage{amsthm}

\usepackage[rightcaption]{sidecap}
\usepackage{changepage}
\usepackage{cancel}
\usepackage{tabularx,colortbl, hhline}
\usepackage{listings}
\usepackage{verbdef}
\usepackage{moreverb}
\usepackage{fancyvrb}
\usepackage{framed}
\usepackage{textcomp}
\usepackage{noindentafter}
\usepackage{atbegshi,ifthen,tikz}
\usepackage[flushleft]{threeparttable}
\usepackage{marginnote}
\usepackage{color}
\usepackage{balance}
\usepackage{flushend}
\usepackage{soul}
\usepackage[noend]{algpseudocode}
\usepackage{mathtools}
\usepackage{hyperxmp}
\usepackage{lipsum}
\usepackage{diagbox}
\usepackage{courier}
\usepackage[english]{babel}
\usepackage{seqsplit}
\hypersetup{
    colorlinks=true,
    linkcolor=blue,
    citecolor=blue,
    filecolor=blue,      
    urlcolor=blue,
}

\usepackage{amsmath,amsfonts,bm}
\usepackage[linesnumbered,ruled,vlined]{algorithm2e}
\usepackage{setspace}

\usepackage{pifont}

\theoremstyle{definition}
\newtheorem{definition}{Definition}[section]

\def\eqref#1{equation~\ref{#1}}

\def\1{\bm{1}}

\DeclareMathAlphabet{\mathsfit}{\encodingdefault}{\sfdefault}{m}{sl}
\SetMathAlphabet{\mathsfit}{bold}{\encodingdefault}{\sfdefault}{bx}{n}

\definecolor{green}{rgb}{0.0, 0.5, 0.0}

\makeatletter

        \newcount\bt@rangea
\newcount\bt@rangeb

\newcommand\btIfInRange[2]{%
	\global\let\bt@inrange\@secondoftwo%
	\edef\bt@rangelist{#2}%
	\foreach \range in \bt@rangelist {%
		\afterassignment\bt@getrangeb%
		\bt@rangea=0\range\relax%
		\pgfmathtruncatemacro\result{ ( #1 >= \bt@rangea) && (#1 <= \bt@rangeb) }%
		\ifnum\result=1\relax%
		\breakforeach%
		\global\let\bt@inrange\@firstoftwo%
		\fi%
	}%
	\bt@inrange%
}

\newcommand\bt@getrangeb{%
	\@ifnextchar\relax%
	{\bt@rangeb=\bt@rangea}%
	{\@getrangeb}%
}

\def\@getrangeb-#1\relax{%
	\ifx\relax#1\relax%
	\bt@rangeb=100000%
	\else%
	\bt@rangeb=#1\relax%
	\fi%
}

\newcommand{\btLstHL}[1]{%
	\btIfInRange{\value{lstnumber}}{#1}%
	{\color{light-gray}}%
	{\def\lst@linebgrd}%
}%

\lst@Key{numbers}{none}{%
    \def\lst@PlaceNumber{\lst@linebgrd}%
    \lstKV@SwitchCases{#1}%
    {none:\\%
     left:\def\lst@PlaceNumber{\llap{\normalfont
        \lst@numberstyle{\thelstnumber}\kern\lst@numbersep}\lst@linebgrd}\\%
     right:\def\lst@PlaceNumber{\rlap{\normalfont
                \kern\linewidth \kern\lst@numbersep
                \lst@numberstyle{\thelstnumber}}\lst@linebgrd}%
    }{\PackageError{Listings}{Numbers #1 unknown}\@ehc}
}

\lst@Key{lbc}{}{%
	\def\lst@linebgrdcolor{#1}%
}
\lst@Key{linebackgroundsep}{0pt}{%
	\def\lst@linebgrdsep{#1}%
}
\lst@Key{linebackgroundwidth}{\linewidth}{%
	\def\lst@linebgrdwidth{#1}%
}
\lst@Key{linebackgroundheight}{\ht\strutbox}{%
	\def\lst@linebgrdheight{#1}%
}
\lst@Key{linebackgrounddepth}{\dp\strutbox}{%
	\def\lst@linebgrddepth{#1}%
}
\lst@Key{linebackgroundcmd}{\color@block}{%
	\def\lst@linebgrdcmd{#1}%
}

\newcommand{\lst@linebgrd}{%
	\ifx\lst@linebgrdcolor\empty\else
	\rlap{%
		\lst@basicstyle
		\color{-.}%
		\lst@linebgrdcolor{%
			\kern-\dimexpr\lst@linebgrdsep\relax%
			\lst@linebgrdcmd{\lst@linebgrdwidth}{\lst@linebgrdheight}{\lst@linebgrddepth}%
		}%
	}%
	\fi
}

\makeatother

\definecolor{light-gray}{gray}{0.85}

\makeatletter
\let\origthelstnumber\thelstnumber

\newcommand*\Suppressnumber{%
	\lst@AddToHook{OnNewLine}{%
		\let\thelstnumber\relax%
		\advance\c@lstnumber-\@ne\relax%
	}%
}

\newcommand*\Reactivatenumber[1]{%
	\setcounter{lstnumber}{\numexpr#1-1\relax}
	\lst@AddToHook{OnNewLine}{%
		\let\thelstnumber\origthelstnumber%
		\refstepcounter{lstnumber}
	}%
}

\makeatother

{\endMakeFramed}

\newcommand\mycomment[1]{}

\SetKwProg{Func}{Function}{}{}
\SetAlFnt{\small\rmfamily}
\SetKw{Continue}{continue}
\SetKw{Break}{break}
\SetKw{Case}{case}
\SetKw{And}{and}
\SetKw{Or}{or}
\SetKw{Not}{not}
\SetKw{In}{in}
\SetKw{Is}{is}
\SetKw{Null}{NULL}
\SetKwInput{Input}{Input}
\SetKwInOut{Output}{Output}

%% file: 0-abstract.tex
As Autonomous driving systems (ADS) have transformed our daily life,
safety of ADS is of growing significance.
While various testing approaches have emerged to enhance the ADS reliability, a crucial gap remains in understanding the accidents causes. Such post-accident analysis is paramount and beneficial for enhancing ADS safety and reliability.
Existing cyber-physical system (CPS) root cause analysis techniques are mainly designed for drones and cannot handle the unique challenges introduced by more complex physical environments and deep learning models deployed in ADS.
In this paper, we address the gap by offering a formal definition of ADS root cause analysis problem and introducing \toolname{}, a novel ADS root cause analysis framework featuring cyber-physical co-mutation.
Our technique uniquely leverages both physical and cyber mutation that can precisely identify the accident-trigger entity and pinpoint the misconfiguration of the target ADS responsible for an accident.
We further design a differential analysis to identify the responsible module to reduce search space for the misconfiguration.
We study 12 categories of ADS accidents and demonstrate the effectiveness and efficiency of \toolname{} in narrowing down search space and pinpointing the misconfiguration. We also show detailed case studies on how the identified misconfiguration helps understand rationale behind accidents.

%% file: 1-introduction.tex
\section{Introduction}

Autonomous driving has achieved remarkable breakthroughs~\cite{lee2017automatic,hakobyan2019high} and becomes closer and closer to our daily life~\cite{howard2014public,yaqoob2019autonomous}. From self-driving cars to delivery robots, autonomous driving techniques are revolutionizing the way we live and work. 
A typical modern autonomous driving system (ADS) employs perception modules to interpret and understand the surrounding environment, prediction and planning algorithms to interact with other vehicles, and controllers to maintain stability and propulsion. These modules can each be incorporated with Deep Learning algorithms, which enhance the ADS's intelligence and adaptability over time.

\begin{figure}
    \centering
    \includegraphics[width=0.95\linewidth]{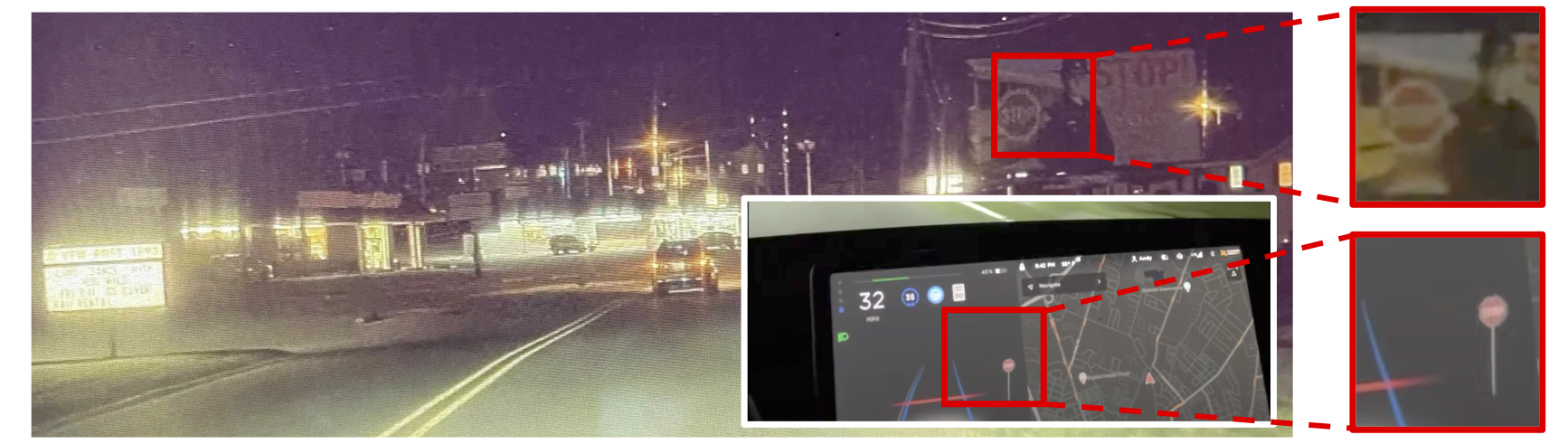}
    \caption{A real-world emergency braking case reported by~\cite{tesla-brake}, as the Tesla autopilot recognizes the stop sign from the road-side billboard. Billboards are common and \textit{benign}, but such scenarios can be \textit{accident-inducing}. \\
    }
    \label{fig:intro-tesla}
    \vspace{-8pt}
\end{figure}

However, as with any complex system, ADS are always prone to errors, which may lead to runtime failures. Due to the inherent uncertainty in Deep Learning models, the complexity of the physical world, and the imprecision in control software~\cite{choi2020cpi}, 
ADS accidents have been witnessed~\cite{news1,news2,news3,news4,news5}, many of them having devastating consequences.
However, the underlying reasons behind these accidents are not readily apparent. For example, as shown in Figure~\ref{fig:intro-tesla}, the Tesla autopilot recognizes the vague stop sign pattern from a road-side billboard, leading to an emergency braking. While the billboard means no harm, such a scenario can be confusing to ADS and potentially induce accidents.
Therefore, post-accident analysis that identifies accident causes is of increasing importance for ADS companies and developers to improve ADS safety and reliability.

Traditionally, post-accident analysis has been performed either in the physical domain, e.g., physical crime scene investigation~\cite{lee2013forensic}, or in the cyber domain, focusing on disclosing trails and provenance of cyber attacks~\cite{zeng2022palantir,mei2022tdlens,kasturi2020tardis}.
However, ADS is essentially a cyber-physical system (CPS) that requires co-analysis of both the cyber and physical worlds. While there are a number of pioneering post-accident analysis techniques in CPS domain, they mainly focus on drone systems rather than ADS. For instance,
MAYDAY~\cite{kim2020mayday} employs program analysis to diagnose accidents caused by controller bugs and mission command bugs using a pre-constructed dependency graph between controllers. RVPlayer~\cite{choi2022rvplayer} decouples aggregated environmental disturbances during logging and applies them to drones for faithful replay.
Although these techniques are effective in their targeted scopes, they can hardly be applied to the ADS domain. 
Firstly, ADS introduces more complex modules such as perception, prediction, and planning modules. MAYDAY relying on the domain specific knowledge of controller programs cannot support other complex modules in ADS.
Furthermore, deep learning models are widely used in ADS, which introduce a significant amount of inherent uncertainty for such traditional program analysis based techniques.
Secondly, ADS operates in much more complicated and interactive physical environments, including a lot of external entities such as other vehicles, pedestrians, traffic lights. 
Thus, it is not straightforward for RVPlayer to decouple and reapply these \textit{indirect} environment influences, and thus the accident replay on ADS is highly challenging.

As far as we know, there is no existing root cause analysis designed for ADS. In this paper, we define ADS root cause analysis as a post-accident analysis to identify the \textit{triggering entity} (e.g., external physical objects) and the \textit{misconfiguration} (e.g., configurable parameters used by ADS) that causes the accident.
We propose \toolname{}, a novel ADS root cause analysis
framework via cyber-physical co-mutation. 
Given an accident, our technique can precisely pinpoint the triggering entity and the misconfiguration of the target ADS. 
Specifically, \toolname first faithfully replay the accident execution inside the simulator, using the recorded locations of ADS and other entities during runtime. 
Then it performs {\it physical mutation} to identify the trigger entity, by finding the minimal environmental entity mutation that suppresses the accident, without changing the ADS’s configuration. Finally, \toolname conducts {\it cyber mutation} to pinpoint the  misconfiguration, by searching for the minimal ADS configuration mutation, without changing the ADS’s trajectory before the accident. 
In practice, cyber mutation proves to be highly time-consuming, primarily due to two reasons. Firstly, the search space is extensive, as exemplified by Baidu Apollo~\cite{baiduapollo}, which encompasses over 1100 configurations. Deciding which subsets of configurations to mutate and determining the magnitude of the mutated values requires considerable effort. Secondly, this search process is not easily parallelizable. Despite the availability of a decent GPU with 8 GB graphical memory, it can only run one simulator and one ADS concurrently. As a result, we further propose a differential analysis algorithm on ADS execution records to reduce the search space for cyber mutation. Details can be found in Section~\ref{sec:design}. Our contributions are summarized as follows.

\vspace{-2pt}
\begin{itemize}
    \item We formally define the problem of ADS root cause analysis and propose \toolname, a novel ADS root cause analysis framework, which incorporates both physical and cyber mutation. These techniques accurately identify the {\it triggering entity} and pinpoint the {\it misconfiguration}, respectively.
    \item We introduce a differential analysis algorithm that effectively reduces the search space for the misconfiguration by comparing two execution records.
    \item We implement a prototype of \toolname{} and evaluate its effectiveness on 12 different types of ADS accidents, including a total of 144 accident cases. Through extensive experiments, we demonstrate that \toolname{} can precisely pinpoint the  misconfiguration responsible for each accident with high efficiency. The code, video demos, and supplementary materials are available at~\cite{rocas-github-repo}. 
    
\end{itemize}

%% file: 2-background.tex
\section{Background}
\label{sec:background}

We introduce the typical ADS architecture 
(Section~\ref{sec:ads-arch}) and ADS communication framework (Section~\ref{sec:ads-commu}).

\subsection{ADS Architecture}\label{sec:ads-arch}
\begin{figure}[t]
    \centering
    \includegraphics[width=.95\linewidth]{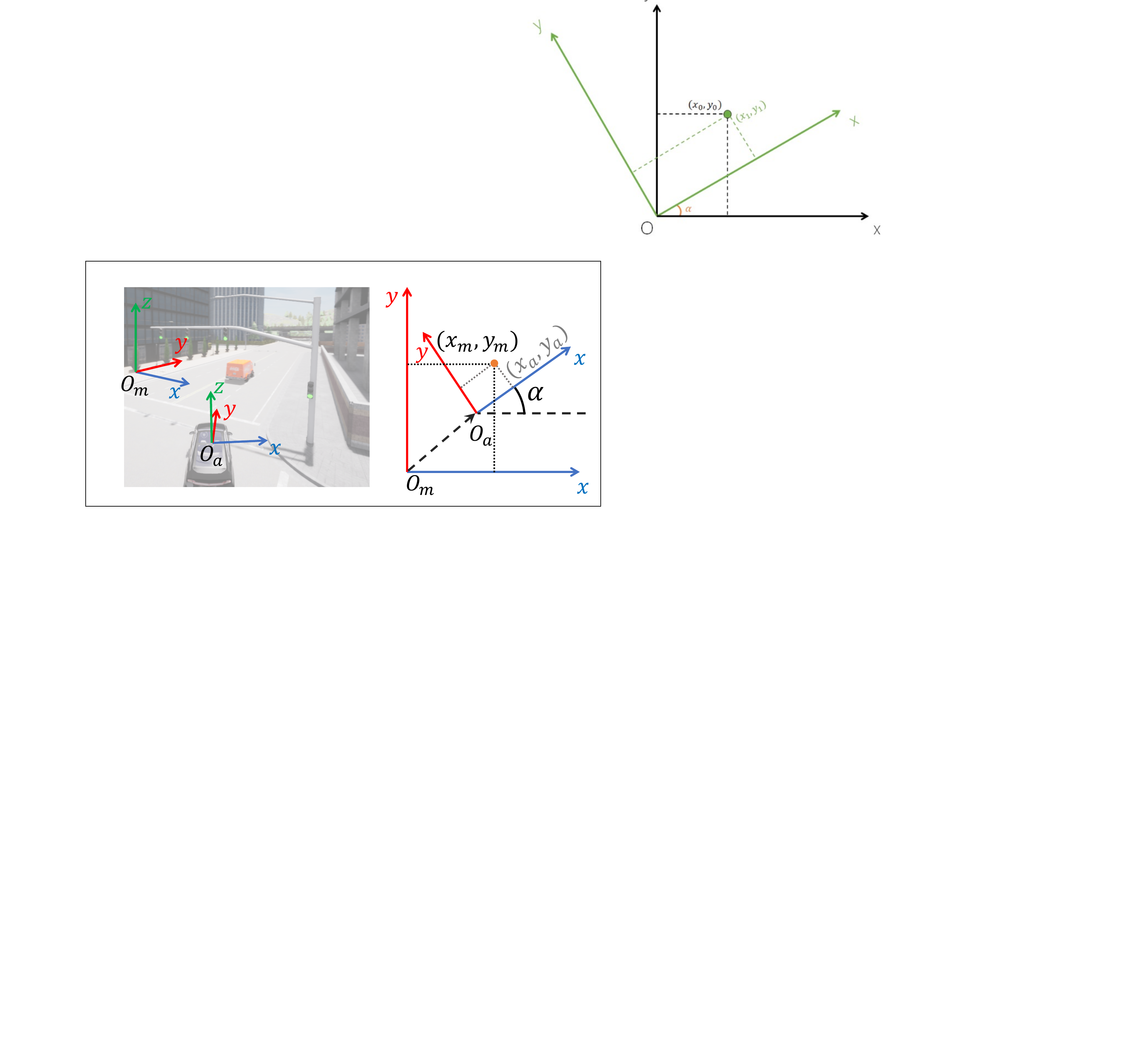}
    \caption{
    A general architecture of ADS with major modules.
    }
    \label{fig:ads_overview}
\end{figure}

ADS architecture contains multiple modules, while the four most closely linked to ADS decision-making are: sensing, perception, prediction, and planning, as shown in Figure~\ref{fig:ads_overview}. 

These modules operate in a cascaded fashion, transforming from the initial hardware sensor input (e.g., video stream and LiDAR point cloud) to the final driving controls (e.g., steering, acceleration, and braking). Specifically,
(1) \textbf{Sensing} utilizes a variety of sensors (cameras, LiDARs, RADARs, etc.) to gather raw environmental data.
(2) \textbf{Perception} interprets this data to understand the environment using deep learning for tasks like traffic sign recognition and object tracking.
(3) \textbf{Prediction} assesses current object statuses to predict future movements and prioritize actions, aiding in accident avoidance.
(4) \textbf{Planning} comprises a global planner for route determination and a local planner for real-time trajectory adjustments based on environmental conditions.
(5) \textbf{Control} executes the driving plan by controlling vehicle movements and adjusting to real-time conditions to ensure adherence to the trajectory.

Each module has hundreds of configurable parameters, such as the obstacle buffer or the reactive braking distance. While these configurations add flexibility to ADS, they simultaneously increase the difficulty of testing and debugging.

\subsection{ADS Communication Framework} \label{sec:ads-commu}
ADS relies on communication frameworks for data transmission among modules, which help decouple the functions of diverse modules and facilitate system development. ADS communication frameworks are typically based on Robot Operating System (ROS). 
ROS employs a publish-and-subscribe mechanism, whereby each \textit{ROS node} can publish data under a specific topic, ensuring that all nodes subscribing to this topic obtain the published data. 
For example, Baidu Apollo~\cite{baiduapollo}, as one of the most popular ADS, utilized ROS as its communication framework in the first few versions and later developed CyberRT\cite{cyber-rt} based on ROS, with optimized efficiency.

%% file: 3-motivation.tex
\section{Motivation} \label{sec:motivation}

This section presents an accident example to illustrate ADS root cause analysis challenges and showcase our technique.

\subsection{Motivating Example}\label{sec:motivating_example}
\noindent \textbf{LAV.}
LAV~\cite{chen2022lav} is an open-source ADS that won the champion of the 2021 CARLA AD Challenge~\cite{carlachallenge}.
Figure~\ref{code:moti} depicts its high-level code logic. It is a representative design of a modularized ADS, following the diagram of \textit{perception-prediction-planning-control}. The function \var{run\_step()} at Line 4 is regularly executed inside the main loop at a frequency of 20Hz. 
At Lines 6-9, the ADS retrieves location information (e.g., \var{gps}), perception raw data (e.g., \var{lidar} and \var{tel\_rgb}), and vehicle states (e.g., \var{speed}) from sensors. The location information \var{gps} is used to determine the road option \var{opt}, e.g., turning right or moving forward (Line 13). 
LiDAR is used to plan ADS's trajectory \var{ego\_plan\_traj} and predict other vehicles' trajectories \var{other\_pred\_traj} (Lines 15-16). 
Telephoto images \var{tel\_rgb} are fed into \var{brake\_model()} to obtain a braking probability \var{pred\_brake} based on traffic lights and hazard conditions (Line 18). The controller \var{pid\_control()} leverages the processed data to produce control commands (i.e., \var{steer}, \var{throt} and \var{brake}) at Lines 22-24. Additionally, post-processings are necessary to ensure adherence to the constraints, such as collision avoidance (function \var{plan\_collide}), hazard stop (\var{BRAKE\_THRESHOLD}), and speed limit (\var{MAX\_SPEED}) at Lines 25-32. 

\input{figtex/moti-code}

\smallskip
\noindent \textbf{Accident Example.} 
Figure~\ref{fig:moti} illustrates the scenario wherein an emergent braking accident is unexpectedly triggered. In this instance, the ADS (shown as the blue car) is following the mission to navigate through the roundabout. However, during its execution, 
the AD vehicle unexpectedly comes to a stop within the roundabout (at the red point). This unforeseen emergency braking within the roundabout could lead to serious rear-end collisions.

\smallskip
\noindent \textbf{Accident Explanation.}
In this example, a red truck ahead ADS triggers a misconfiguration \var{BRAKE\_THRESHOLD} and leads to an accident.
As shown at Line 27 in Figure~\ref{code:moti}, the variable \var{pred\_brake} is compared with a fixed threshold \var{BRAKE\_THRESHOLD}, whose value is set to 0.1 at Line 2, to determine if a brake is necessary.
\var{pred\_brake} means the probability of braking, computed by a DL model  \var{brake\_model} (Line 18) which takes as input the images from ADS’s camera with telephoto lens . 
Figure~\ref{fig:moti-front-camera}a shows the captured images when the AD vehicle stops within the roundabout.
In normal cases, without a red light, the \var{brake\_model()} outputs a value  (i.e., \var{pred\_brake}) close to zero, which is smaller than the \var{BRAKE\_THRESHOLD}. 
However, in this example, there is a red struck positioned at this particular angle, which shares partial feature with a red traffic light and makes the variable \var{pred\_brake} fluctuate around 0.2. Note that this is still a relatively small braking probability. However, after comparison with the fixed threshold, the variable \var{brake} is set to 1 at Line 28 and passed to the controller at Line 34, resulting in an emergency braking.

\subsection{Limitation of Existing Works}

After an accident happens, post-accident analysis must be conducted to identify the underlying root causes, including both the external triggering entities (e.g., the red truck in Figure~\ref{fig:moti}) and the internal  misconfigurations (e.g., the variable \var{pred\_bake}), and improve the reliability of the ADS. 

Unlike traditional software root cause analysis, ADS is a cyber-physical system (CPS), which requires the joint analysis of both the cyber and physical domains.
Existing CPS post-accident analysis frameworks mainly focus on drone systems. There are two main types of techniques, i.e., program analysis based and what-if reasoning based.
These post-accident analysis for drone systems usually focus on the controller components and primarily address simple environmental factors such as weather conditions (e.g., wind gust).
However, ADS has a much more complex architecture, equipped with sophisticated modules such as perception, prediction, and planning modules, and also with deep learning models. 
In addition, ADS operates in much more complicated physical environments, including external entities such as vehicles, pedestrians, and traffic lights. These factors make it difficult to directly apply existing methods for drone systems to ADS scenarios. 

\noindent
\textbf{Program analysis based method.}
MAYDAY~\cite{kim2020mayday} utilizes program analysis to help diagnose accidents caused by controller bugs and mission command bugs. It leverages a pre-constructed dependency graph between controllers to find the state deviation. 
Thus, MAYDAY can only identify a potentially problematic basic block in control program, without pinpointing a specific line of code or a configuration. Also, it is based on the domain specific knowledge of controller programs and cannot support other complex modules in ADS, such as the perception, prediction, and planning.
Furthermore, deep learning models are widely used in ADS, which introduce a significant amount of inherent uncertainty for traditional program analysis based techniques. 

\noindent
\textbf{What-if reasoning based method.}
Due to the limitation of program analysis based techniques, some works investigate the accident root causes by storing the environmental disturbances and replaying the execution. 
RVPlayer~\cite{choi2022rvplayer} decouples the aggregated environmental disturbances during logging and reapplies this disturbance on drones to enable faithful replay. 
However, ADS operates in much more complicated and interactive physical environments, including a lot of external entities such as other vehicles, pedestrians, traffic lights. 
Different from drone systems that are primarily {\it reactive} to environmental disturbances, ADS is required to be {\it proactive} to handle these intricate situations. 
Thus, it is not straightforward to decouple and reapply these indirect environment influences, and the accident replay on ADS is highly challenging.

\subsection{Our Approach}

In this paper, we propose an ADS root cause analysis featuring {\it cyber-physical co-mutation} to identify the accident causes. 
To apply the mutation, we first need to reconstruct the accident in a simulator (we call it the accident execution). We instrument the ADS to record its own locations from GPS and the bounding box (including relative locations with respect to ADS) of other vehicles. Given the map file, all the locations can be transformed into simulator locations, enabling the simulator to replay the accident execution.

After replaying the accident execution in the simulator, we apply two steps of mutation to identify the triggering entities and  misconfigurations, separately.
First, we keep all ADS configurations (i.e., the cyber space) unchanged, only mutate the physical conditions, replay the execution in the simulator, and observe if the accident still occurs. 
We call this process as \textit{physical mutation}, aiming to find the minimal mutation (in physical space) to suppress the accident.  
The example in Section~\ref{sec:motivating_example} demonstrates an emergency braking accident. It is possible that a specific physical condition may fool deep learning models used by ADS and cause them to mistake a red light or an obstacle on the path, resulting in an emergency braking decision.
Figure~\ref{fig:moti-front-camera}a displays the images captured by the ADS’s camera at the emergency braking point within the roundabout. If we disable the red truck appearing in the middle region of the cameras (as shown in Figure~\ref{fig:moti-front-camera}b), while keeping all configurations and other physical conditions unchanged, the subject vehicle will not stop within the roundabout during replay. Thus, we know that the red truck is the triggering entity that leads to the accident.

In the second mutation, we search for the misconfiguration. Unlike physical mutation, in this step, we only mutate the configurations inside ADS, while keeping the physical space. The mutation in this process is referred to as \textit{cyber mutation}. This mutation aims to find the minimal mutation (in cyber space) to suppress the accident and outputs the misconfigurations that lead to the accident.
For example, there are two configurations shown in Figure~\ref{code:moti}, i.e., \var{MAX\_SPEED} and \var{BRAKE\_THRESHOLD} (Line 1-2). When we increase the value of \var{BRAKE\_THRESHOLD} to 0.5, the accident is suppressed during the replay, and the subject vehicle can successfully pass through the roundabout. However, if we mutate another configuration, such as \var{MAX\_SPEED} (e.g., set its value to 50), the accident still occurs. This reveals that \var{BRAKE\_THRESHOLD} is highly likely to be the misconfiguration and helps us identify Line 27, which is responsible for the accident. 

However, a challenge is that cyber mutation can be very time-consuming due to the large search space of configurations. To reduce search space, we identify the initial deviating module that is likely to cause the accident before searching for misconfigurations in cyber mutation.
Specifically, we record {\it the channel messages communicated among modules} to represent an execution record. Then, we conduct differential analysis on the reference execution records (obtained from physical mutation) and the accident execution records, to facilitate the initial deviating module.

To summarize, we propose \toolname, an ADS root cause analysis framework for post-accident analysis. The overall system design of \toolname is shown in Figure~\ref{fig:overview}. 

In Phase-I, given an accident, \toolname replays the accident execution in simulation. Then in Phase-II, \toolname conducts physical mutation on the accident execution, producing the triggering entity and an accident-free reference execution. In Phase-III, by doing differential analysis on the two execution records from accident execution and reference execution, \toolname identifies the initial deviating module in order to reduce search space for Phase-IV. Finally in Phase-IV, \toolname runs cyber mutation only within the identified module and outputs the misconfiguration. More details are elaborated in Section~\ref{sec:design}.

%% file: figtex/moti-code.tex
\begin{figure*}[t!]
\setlength{\tabcolsep}{8pt}
\begin{tabular}{>{\RaggedLeft}p{0.39\linewidth}p{0.58\linewidth}}
\vspace{0pt} %
\begin{minipage}{0.9\linewidth}
\centering
\begin{lstlisting}[basicstyle=\footnotesize, boxpos=t, lbc = {\btLstHL{2, 18, 27}}]
const MAX_SPEED = 35 // km/h
const BRAKE_THRESHOLD = 0.1
MISSION = read_config_file()
def @run_step@(input_data) {
    // Get sensor data and vehicle states
    gps      = input_data.get("'GPS'")
    lidar = input_data.get("'LIDAR'")
    tel_rgb  = input_data.get("'TEL_RGB'")
    speed      = input_data.get("'ego_speed'") // m/s
    ...
    // Get road option from global planner, 
    // e.g., RIGHT/LEFT/FORWARD/STOP. 
    opt = waypointer(MISSION, gps)
    // Motion prediction & local planning
    ego_plan_traj, other_pred_traj = \
                        infer_model(lidar, opt)
    // Brake prediction from telephoto lens images
    pred_brake = brake_model(tel_rgb)
    ...
    // Control command
    steer, throt, brake = 0, 0, 0
    if not has_none_val(ego_plan_traj):
        steer, throt, brake = \
            pid_control(ego_plan_traj, speed, opt)
    collide_flag = plan_collide( \
        ego_plan_traj, other_pred_traj)
    if pred_brake > BRAKE_THRESHOLD:
        throt, brake = 0, 1
    elif collide_flag:
        throt, brake = 0, 1
    if speed * 3.6 > MAX_SPEED:
        throt = 0
    ...
    return steer, throt, brake
}
\end{lstlisting}
\caption{High-level Code Logic of LAV~\cite{chen2022lav}}\label{code:moti}
\end{minipage}
&
\vspace{0pt} %
\begin{minipage}{0.9\linewidth}
    \includegraphics[width=0.8\linewidth]{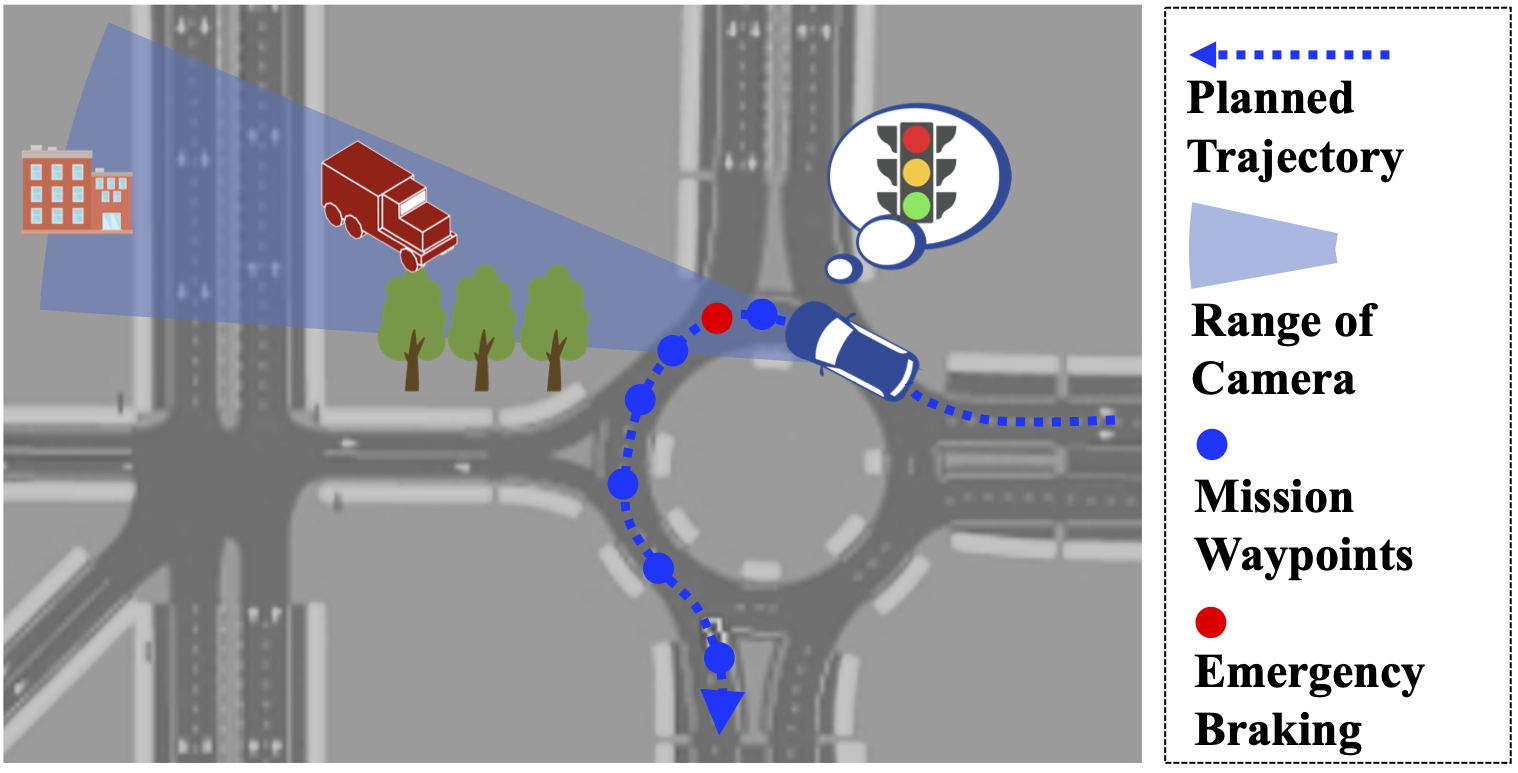}
    \centering
    \caption{An accident example for LAV. The vehicle unexpectedly stops within the roundabout, triggered by the red truck ahead.}
    \label{fig:moti}
\end{minipage}
~
\begin{minipage}{0.9\linewidth}
    \vspace{11pt}
    \begin{subfigure}{\linewidth}
        \includegraphics[width=0.85\linewidth]{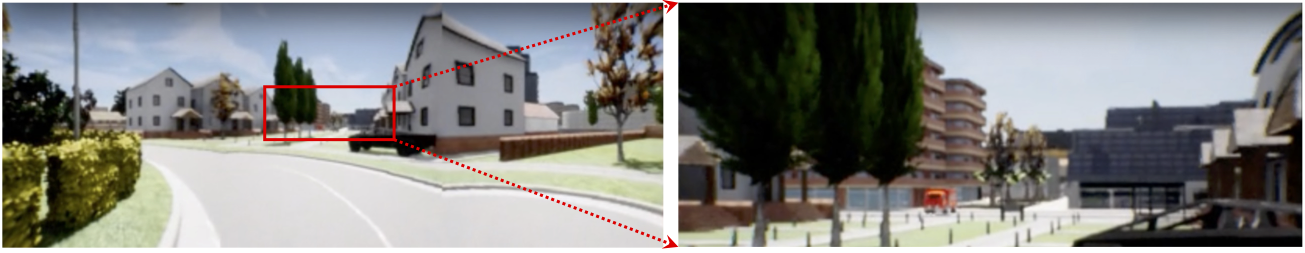}
        \centering
        \caption{Emergency Braking.}
        \label{fig:moti-before}
    \end{subfigure}
    ~
    \begin{subfigure}{\linewidth}
        \includegraphics[width=0.85\linewidth]{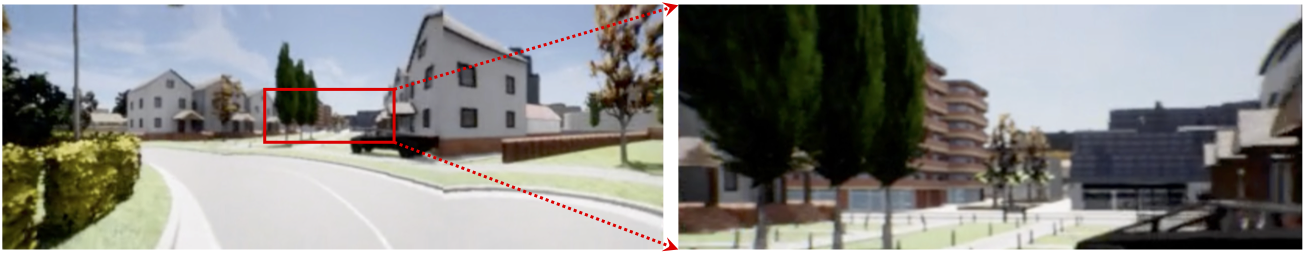}
        \centering
        \caption{No Emergency Braking.} 
        \phantomcaption
        \label{fig:moti-after}
    \end{subfigure}
    \caption{ 
        Subfig (a) shows the image captured by ADS's cameras when it stops within the roundabout. Subfig (b) removes the red truck in the middle region and prevents the accident. %
    } \label{fig:moti-front-camera}
\end{minipage}
\end{tabular}
\vspace{-10pt}
\end{figure*}

%% file: 4-definition.tex
\section{\large Definition of ADS Root Cause Analysis}
We formally define ADS execution and accident root cause analysis.

\vspace{-5pt}
\subsection{ADS Execution}
\label{sec:ads-exec}
An execution takes as input a configurable ADS and physical environment (e.g., map, weather, and other vehicles).
The execution records how the ADS reacts to the environment and reports whether an accident happens during the execution.

\noindent {\bf Configurable ADS.} Formally, we denote the type of the cyber configuration as 
$\mathcal{C} \coloneqq \mathcal{C}_1 \times \mathcal{C}_2 \times \cdots \times \mathcal{C}_r,$
where $\mathcal{C}_i  \coloneqq \mathbb{R}^{\mu_i}$ denotes the type of the configuration for module $i$.

\noindent {\bf Physical environment.}
Physical environment consists of two parts: the actor entities (e.g., moving vehicles) and the map entities (e.g., traffic cones, traffic lights, and buildings).  
Formally, we use $\mathcal{A}\coloneqq  \mathcal{A}_1 \times \mathcal{A}_2 \times \cdots \times \mathcal{A}_j$ and $\mathcal{M}\coloneqq \mathcal{M}_1 \times \mathcal{M}_2 \times \cdots \times \mathcal{M}_k,$ to denote the type of the states for actor entities and map entities, respectively.
Specifically, supposing that each entity has $\sigma$ properties (e.g., location, orientation, velocity), $\mathcal{A}_i \coloneqq \mathbb{R}^{\sigma}$ and $\mathcal{M}_i \coloneqq \mathbb{R}^{\sigma}$ denotes the type of the states the $i$th actor entity and that for the $i$th map entity, respectively.

\noindent \textbf{ADS execution.}  One complexity of ADS execution is that properties of an actor entity may change over time.
We thus utilize $A_i@t$, where $A_i \in \mathcal{A}_i$, to represent the state of the $i$-th actor entity at timestamp $t$. Moreover, we adopt $A_i@[t_1, t_2]$ to denote a sequence of states of the $i$-th actor entity during the time span $[t_1, t_2]$.
For simplicity, we use ${A}@t$, where $A_i \in \mathcal{A}$, to denote the state of all actor entities at the timestamp $t$ and ${A}@[t_1,t_2]$ the state sequences of all actor entities during the time span $[t_1, t_2]$.

Formally, we use $\mathcal{E}$ to denote the type of an execution: $\mathcal{E}: \mathcal{C} \times \mathcal{M} \times \mathcal{A} \rightarrow \text{List}[\mathcal{A}] \times \mathbb{B}$.
Intuitively, an execution takes as input the configuration of the ADS, the states of the map entities and actor entities, and outputs the states of every actor entity at all timestamps. Moreover, it also outputs a boolean value indicating 
whether an accident happens (i.e., collision or emergency braking).

To facilitate discussion, we use $C$ to denote a concrete instance of the cyber configuration; 
Similarly, $A$ for actor entities, $M$ for map entities and $E$ for an execution.
Note that we assume the first element in $A$ always denotes the AD vehicle, namely $A_0$.

\subsection{ADS Accident Root Cause Analysis} \label{sec:acc-inv}
Root cause analysis for ADS accident aims to identify the triggering entities and the misconfigurations.
We formally define an accident execution as
$
    E\Big( C, M, A@t_0 \Big)
    \rightarrow \langle  A@[t_0,t_0+T], True \rangle.
$
We further assume the accident happens after a time period of $d$ after starting.

\smallskip
\noindent{\bf Triggering entities identification.} 
We define the triggering entities as a set of actor and map entities such that minimal changes to their states (noted as $\Delta M\in\mathcal{M}$ and $\Delta A \in \mathcal{A}$)
would:\\
\noindent (1) {\it suppress the accident.} 
Formally, 
    \vspace{-2pt}
    \begin{align*}
    E\Big( C, M \oplus \Delta M, A@t_0 \oplus \Delta A\Big)
    \rightarrow \langle  A@[t_1,t_1+T], False \rangle,
    \end{align*}
where $\oplus$ denotes applying changes to the states values.\\
\noindent (2) and {\it introduce limited changes to the actor entities before entering the accident scene.}
Formally,
    \vspace{-2pt}
    \begin{align*}
    \Sigma_{t \in [0,d-\delta]}
    \Big (\big|\big| A@(t_0+t) - A@(t_1+t) \big|\big| \Big) < \epsilon,
    \end{align*}
where $\delta$ denotes a short time period before the accident occurs, and
$\lVert \cdot \rVert$ denotes $L^2$ norm.

\smallskip
\noindent{\bf Misconfiguration identification.} After finding triggering entities, we further search for misconfigurations.
Formally, we search for
minimal changes of configurations, noted as $\Delta C \in \mathcal{C}$, such that~\looseness=-1

\noindent (1) {\it the accident is suppressed.} 
Formally, 
    \vspace{-2pt}
    \begin{align*}
    E\Big( C \oplus \Delta C, M, A@t_0 \Big)
    \Rightarrow \langle  A@[t_2,t_2+T], False \rangle,
    \end{align*}
    \vspace{-1pt}
where $\oplus$ denotes applying changes to the original configurations.
 
\noindent (2) {\it the AD vehicle does not change its behavior before entering the accident scene.} 
Formally,
    \vspace{-2pt}
    \begin{align*}
    \Sigma_{t\in [0,d-\delta]} 
    \Big( \big|\big| A_0@(t_0+t) - A_0@(t_2+t) \big|\big| \Big) < \epsilon,
    \end{align*}
where $\delta$ denotes a short time period before the accident occurs and
$\lVert \cdot \rVert$ denotes the $L^2$ norm.
This requirement ensures the AD vehicle encounters a similar scenario before and after the configuration changes. Otherwise, one can always 
freeze the AD vehicle to suppress the accident, which is meaningless for the investigation.

%% file: 5-design.tex
\begin{table}
     \centering
    \setlength{\tabcolsep}{1.5pt}
    \footnotesize
    \captionof{table}{Physical Mutation Space.}
    \label{tab:phys-mut-space}
    \begin{tabular}{cccc}
        \toprule
        Domain           & Category     & Configuration     & Data Type                 \\
        \midrule
        \multirow{8.5}{*}{\makecell[c]{Actor\\Entities}} 
            & \multirow{5}{*}{Vehicles} & Model     & \{Sedan, Truck, ...\}     \\
            &                           & Color     & \{Red, Blue, Black, ...\} \\
            &                           & Location  & [x y z]                   \\
            &                           & Rotation  & [yaw pitch roll]          \\
            &                           & Speed     & S (m/s)               \\
            \cmidrule{2-4}
            & \multirow{3}{*}{\makecell[c]{Cyclists/\\Pedestrians}} & Location  & [x y z]           \\
            &                                                       & Rotation  & [yaw pitch roll]  \\
            &                                                       & Speed     & S (m/s)      \\
        \midrule
        \multirow{12.5}{*}{\makecell[c]{Map\\Entites}}   
            & \multirow{2}{*}{\makecell[c]{Traffic Cones/\\Boxes}} & Location   & [x y z]           \\
            &                                                      & Rotation   & [yaw pitch roll]  \\
            \cmidrule{2-4}
            & Buildings                         & Enable        & \{True, False\}   \\
            \cmidrule{2-4}
            & Vegetations                       & Enable        & \{True, False\}   \\
            \cmidrule{2-4}
            & \multirow{2}{*}{Traffic Lights}   & Policy        & \{Red, Yellow, Green\}    \\
            &                                   & Enable        & \{True, False\}           \\
            \cmidrule{2-4}
            & \multirow{5}{*}{Weather}          & Cloudiness            & [0, 100]                \\
            &                                   & Precipitation         & [0, 100]                \\
            &                                   & Sun Azimuth Angle     & [0, 360] (deg)          \\
            &                                   & Sun Altitute Angle    & [-90, 90] (deg)         \\
            &                                   & Fog Density           & [0, 100]               \\
        \bottomrule
    \end{tabular}
\end{table}

\section{System Design}\label{sec:design}

In this section, we discuss the design details of \toolname. 
First, \toolname replays the accident execution by applying coordinate transformation on execution records.
Then it conducts physical mutation (Section~\ref{sec:phy-mut}) to find the accident-triggering entities. Moreover, the physical mutation further yields a reference execution without accident.
To search for the misconfigurations,
\toolname first pinpoints the initial deviating module %
(Section~\ref{sec:design-id-module}) to reduce search space via differential analysis between the accident execution record and the reference execution record.
Finally \toolname performs cyber mutation to identify the misconfigurations (Section~\ref{sec:cyber-mut}).

\input{tables/trans_n_physmut}
\begin{figure*}[t]
    \centering
    \includegraphics[width=.95\linewidth]{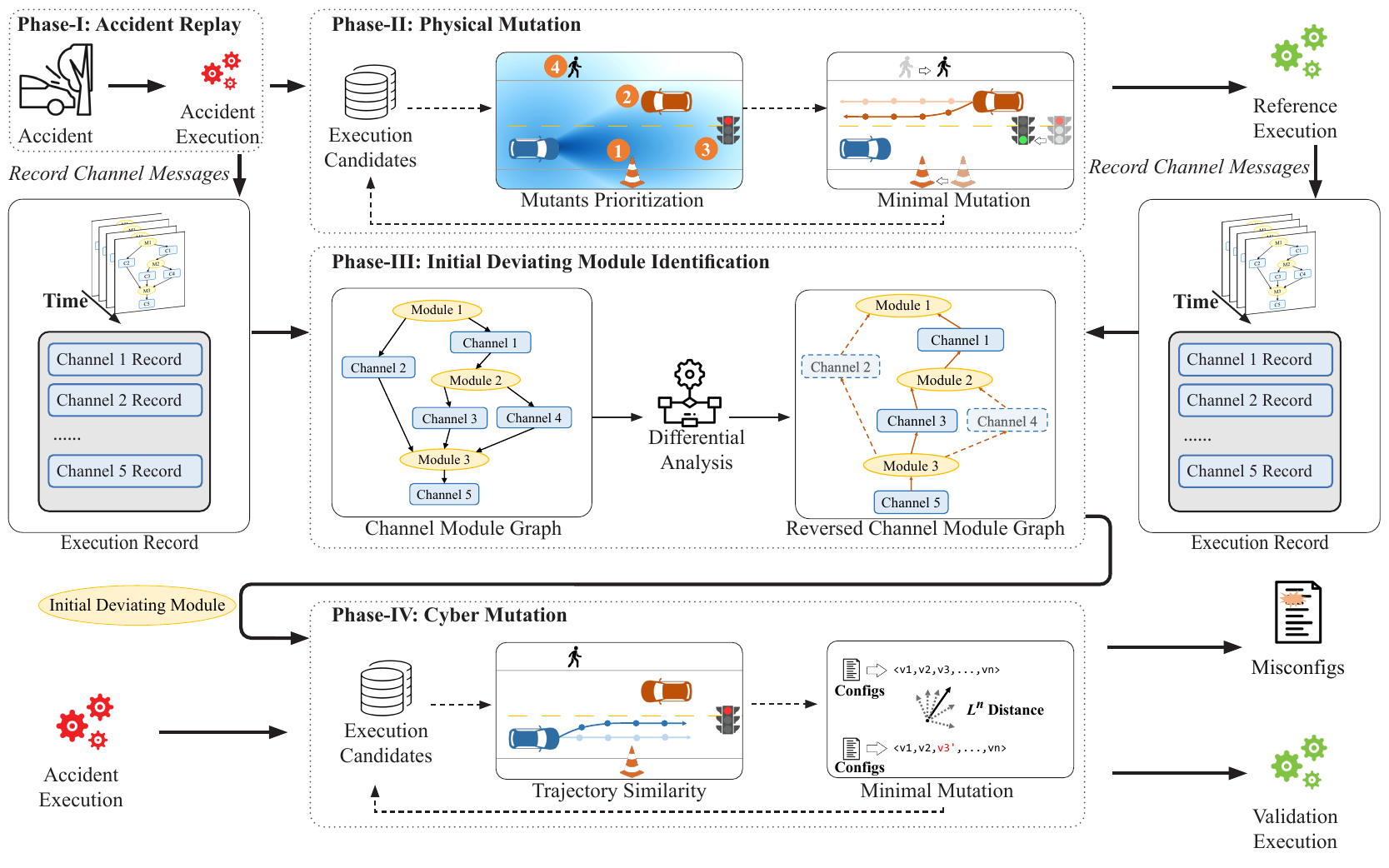}
    \vspace{-10pt}
    \caption{System design of \toolname
    \vspace{-10pt}
    }
    \label{fig:overview}
\end{figure*}

\subsection{Physical Mutation}
\label{sec:phy-mut}
Physical mutation serves two purposes: (1) it helps produce the accident root cause results by finding the accident-triggering entities; (2) it outputs a reference execution that is similar to the original execution but accident-free. An accident-free reference execution indicates that the accident could have been avoided in a similar scenario. The differential analysis in the later stage will compare these two execution records to facilitate localizing the module that has accident-inducing behaviors.

To search for an accident-free execution, the physical mutation stage mutates the surrounding entities (i.e., $A$, $M$ defined in Section~\ref{sec:ads-exec}) without modifying the configuration of the AD vehicle. 
The entity space, which is also \toolname's mutation space, is shown in Table~\ref{tab:phys-mut-space}.

We formulate the physical mutation stage as a Multi-object Optimization Problem (MOP)~\cite{mop}. The MOP algorithm takes as input a set of objective functions, a set of constraints, and a set of variables from the solution domain. It employs a genetic algorithm and outputs an optimal set of variables with the minimal objective values and meanwhile satisfy the constraints.
In this stage, we leverage the MOP algorithm to search for a set of properties for the surrounding entities with minimal changes to the original execution and eliminate the accident.

Our fitness function requires that 
(a) the replay execution yields false. (i.e., the accident is suppressed); 
(b) the mutated entity is highly related to the triggering condition; 
(c) the replay scenario should be similar to the original one; 
(d) the set of mutated physical conditions is small; and 
(e) their value changes are small.

Following the notations discussed in Section~\ref{sec:ads-exec}, $E(\cdot)$ returns whether the accident happens. Given an initial physical condition $p_0 \in \mathcal{C} \times \mathcal{A} \times \mathcal{M}$ such that $E(p_0)=1$, finding the triggering entity (i.e., minimum mutation) can be formulated as:
\begin{equation}
\label{eq:phys-mop}
\begin{split}
    \textbf{minimize } \mathcal{F}(p) &= \{ f_1(p), f_2(p), f_3(p) \}, \\ 
    \textbf{subject to } \mathcal{G}(p) &= \{ E(p)=0 \}.
\end{split}
\end{equation}
\indent $\mathcal{F}$ represents the function we aim to minimize, while $\mathcal{G}$ denotes the constraint that must be satisfied.
Specifically, $f_1(\cdot)$ defined in Eq.~\ref{eq:f1} quantifies the suspiciousness of the mutated entity and satisfies the requirement (b);
$dist_i$ and $\theta_i$ refer to the distance and angle from mutated entity $i$ to the ADS; $K$ is a parameter that balances between distance and angle. Intuitively, the entity that is close to and in front of the ADS are more likely to be accident-inducing than those far away or behind the ADS. As visualized in the ``Mutants Prioritization'' figure in Figure~\ref{fig:overview} ``Phase-I'' block, darker blue denotes higher relevance, while white denotes not relevant. 
\begin{equation}\label{eq:f1}
\begin{split}
    f_1(p) = - \sum_{i \in p-p_0} \Big( K / dist_i + cos(\theta_i) \Big)
\end{split}
\end{equation}
\indent The $f_2(\cdot)$ satisfies the requirement (c) that the replay scenario should be similar to the original one. We quantify the scenario similarity using the trajectory similarity (computed by MSE, mean squared error) of all actor entities. $Traj_{p}(j)$ denotes the trajectory of actor entity $j$ under the physical condition $p$. This objective function is necessary, otherwise we can just set all actor entities' speed to zero and the accident can be suppressed.
\begin{equation}\label{eq:f2}
    f_2(p) = - \sum_{j \in A} Sim\Big(Traj_{p_0}(j), Traj_p(j)\Big)
\end{equation}
\indent Function $f_3$ defines the distance of an offspring from the initial physical configuration $p_0$, satisfying the requirements (d) and (e).
The first term measures that the set of mutated physical conditions, which should be small. The second and the third terms quantify that the value changes of mutated physical conditions should be small. We use $L_1$-Norm to compute the edit distance of enumerate and boolean type configurations and use $L_2$-Norm for other types.
\begin{equation}\label{eq:f3}
\begin{split}
    f_3(p) = \#(p-p_0) + \sum_{k \in Enum} L_1 (k_0, k) + \sum_{k \notin Enum} L_2 (k_0, k)
\end{split}
\end{equation}

\subsection{Initial Deviating Module Identification}\label{sec:design-id-module}

After finding an accident-free reference execution, \toolname differentiates 
the reference execution and the original execution to localize the module that potentially has misconfigurations. %
There are two major challenges in differentiating two executions records. 
First, due to the noise in the physical world, even two executions with the same setup may have differences. It is thus important to distinguish the differences that lead to different behaviors of the ADS and the differences that are caused by noise.
Moreover, since the reference execution and the original execution have slightly different setup, \toolname may observe differences in multiple modules of the system. For example, if a position of an actor car is slightly changed, both vision module and planning module may output different results. It is thus essential to locate the accident-inducing differences.

We propose to represent an ADS as a channel module graph (CMG), and use the messages in channels to represent an execution. \toolname differentiates the messages to locate the problematic module and traverse the CMG to identify the root cause.

\subsubsection{Graph representation for ADS}
\label{sec:graph-repr}

Modules in an ADS use channels to communicate with each other. A channel represents a segment of shared memory within the ADS. %
A module reads messages from other modules via channels and writes its computational outcomes to output channels. %
Different modules run in parallel to achieve better real-time performance. 
Traditional program representations, like control flow graph or data dependency graph, cannot handle the \textit{temporal} (e.g., variable values are updated tens or hundreds of times at each second) %
and \textit{heterogeneous} (e.g., involve multiple modules and deep learning models) features of ADS. We thus propose channel module graph (CMG) defined in Definition~\ref{def:cm-graph} to represent ADS programs.
We use an example from Apollo~\cite{baiduapollo} to illustrate how \toolname leverages CMG to represent ADS structure.

\begin{definition}[Channel Module Graph]\label{def:cm-graph}
Given an ADS, we use a directed bipartite graph, \textit{Channel Module Graph (CMG)}, to represent ADS topology structure, denoted as $G=\langle V_{C}, V_{M}, E \rangle$. \\
(1) Two disjoint vertices set $V_C$ and $V_M$ represent \textit{channels} and \textit{modules} of ADS, respectively. \\
(2) $e=\langle v_M, v_C \rangle \in E, v_M \in V_M, v_C \in V_C \iff v_M$ \textit{writes to} $v_C$.\\
(3) $e=\langle v_C, v_M \rangle \in E, v_M \in V_M, v_C \in V_C \iff v_M$ \textit{reads from} $v_C$.
\end{definition}

\input{figtex/msg-reverse-graph}

A concrete example is shown in Figure~\ref{fig:running-ex}a. Note that the CMGs of the original execution and the reference execution are the same since the reference execution only mutates the surrounding environments without altering ADS code.

\subsubsection{Message as execution records} \label{sec:msg-record}
We use messages that communicated in channels as the records of ADS execution, considering its two advantages: (1) the messages naturally capture the temporal feature of ADS since a typical message contains the timestamp. (2) the message is agnostic of the possibly heterogeneous implementation of multiple modules.\\
\indent Figure~\ref{fig:running-ex}b shows (simplified) message definition for the \textit{planning} channel. Each planning channel message consists of (1) a \var{lane\_id} field, denoting the lane that AD vehicle should be on; (2) a \var{traj\_point} field, containing a sequence of trajectory points that represents the expected position of the AD vehicle.
Due to the nested structure of messages, one piece of message can be easily transformed into its corresponding tree representation (Figure~\ref{fig:running-ex}c).

\subsubsection{Differential analysis on execution records}\label{sec:diff-exec}
Recall that we now have two records, one obtained from the replay accident execution, the other from the accident-free reference execution. We conduct differential analysis on these two execution records to locate the responsible module for the accident occurrence. \\
\indent We introduce the metric {\it Message Difference Ratio} ($\mathtt{MDR}$) to quantify the difference between two messages. Given two execution records, \toolname computes \metric{MDR} for each channel at each timestamp.
\begin{definition}[Message Difference Ratio]
Given two messages $m_1$, $m_2$ (with tree representations $tr_1$, $tr_2$), their \textit{message difference ratio} (\metric{MDR}) are computed as follows:
\begin{align*}\label{eqt:tree-diff} 
\footnotesize
\begin{split}
 \mathtt{MDR}(tr_1, tr_2) = 
  & \begin{cases}
    1\textnormal{\bf\ if } tr_1 \neq tr_2 \textnormal{\bf\ else } 0, 
    & |tr_1.chd| = |tr_2.chd| = 0 \\
    1, 
    & |tr_1.chd| \neq |tr_2.chd| \\
    \dfrac{\sum\limits^{|tr_1.chd|}_{i=0} \mathtt{MDR}(tr_1.chd[i], tr_2.chd[i])}{|tr_1.chd|},  
    & Otherwise 
  \end{cases}
\end{split}
\end{align*}
\end{definition} 
Intuitively, \metric{MDR} takes as input two trees, and outputs a difference score from 0 (same) to 1 (different). 
The first case means both input trees are leafs. The two leafs are then directly compared. 
The second case indicates that two input trees have different numbers of children. We simply consider them as different.
For two non-leaf tree nodes with the same number of children, we recursively compare each child, and use the average \metric{MDR} scores of all children as the \metric{MDR} score of the two input trees.\\
\indent As the output of differential analysis, for each channel, we obtain a series of \metric{MDR} at each timestamp. These \metric{MDR} series are further leveraged to pinpoint the initial deviating module.

\subsubsection{Pinpoint initial deviating module}\label{sec:pinpoint-module}

Our insight is that: (1) {\it the timestamp that \var{control} channel has a sudden change of \metric{MDR} is the occurrence time of the accident.} (2) {\it the channels that has sudden changes of \metric{MDR} before the accident occurrence time are likely to be responsible for the accident.}\\
\indent \toolname leverages the \func{PELT}~\cite{pelt} change point detection algorithm to detect when \metric{MDR} changes significantly for each channel.
The \func{PELT} algorithm detects significant change points from a sequence of values. 
The rationale is that the first few seconds of compared executions are expected to be similar, %
thus \metric{MDR} in the first few seconds can be considered as the baseline noise of a channel.
A significant change on \metric{MDR} value indicates that the difference among channel messages significantly goes beyond noise. \\
\input{figtex/init-module}
\indent Detailed pinpoint algorithm is shown in Alg.~\ref{alg:init-module}. It takes 2 inputs, the CMG $G$ and the \metric{MDR} series of all channels {\it channel2mdr} obtained from differential analysis in Section~\ref{sec:diff-exec}. Output is the identified initial deviating module that are responsible for the accident.\\
\indent Specifically, given {\it channel2mdr}, Alg.~\ref{alg:init-module} first uses \func{PELT} to compute a list of changing time points for each channel (Line 3).
Then $t^*$ is used to store the earliest changing time points for each node (Line 4-8). After that, we remove the edge $(u,v)$ from $G$ if the sudden \metric{MDR} change of $u$ does not lead to the \metric{MDR} change of $v$ (Line 9-11). \toolname use $\delta=3$ because 3 seconds is a common reaction time in driving.
Finally, starting from $ctrlnode$, the algorithm traverses from the reversed CMG $revG$, shown in Fig.~\ref{fig:causality-graph}, and stores all reachable nodes in set $D$ (Line 12-14). The algorithm returns the module in $D$ that has the earliest time changing point of \metric{MDR}, namely the initial deviating module.

\subsection{Cyber Mutation}\label{sec:cyber-mut}

After Alg.~\ref{alg:init-module}, we have obtained the initial deviating module. 
As the final phase to localize the accident-inducing cyber parameter setting, \toolname aims to find a minimal mutation on the parameter space within the identified module so that the original accident can be suppressed. We use a similar approach as Section~\ref{sec:phy-mut}. Note that the difference from Section~\ref{sec:phy-mut} is that \toolname only mutates the internal cyber parameters in this phase.\\
\indent The main motivation for modifying multiple configurations is that modifying only one configuration per iteration is much slower. In practice, the majority of the thousands of ADS configurations are irrelevant to the accident. Modifying multiple configurations simultaneously is more likely to hit the responsible configurations in a shorter time.\\
\indent Our fitness function requires 
(1) the replay execution yields false. (i.e., accident suppressed); 
(2) the ADS's trajectory after mutation should be similar to the original one; 
(3) the set of mutated physical configurations is small; and 
(4) their value changes are small.\\
\indent Given an initial cyber configuration $c_0 \in C$ such that $E(c_0)=1$, finding the misconfiguration (i.e., minimum mutation) can be formulated as:
\begin{equation}
\begin{split}
    \textbf{minimize }   \mathcal{H}(c) &= \{ h_1(c), h_2(c) \}, \\
    \textbf{subject to }  \mathcal{K}(c) &= \{ E(c)=0 \}
\end{split}
\end{equation}\label{eq:cyber-mop}
\indent $\mathcal{H}$ represents the function we aim to minimize, while $\mathcal{K}$ denotes the constraint that must be satisfied. $h_1(\cdot)$ defined in Eq.~\ref{eq:h1} measures the trajectory similarity (computed by MSE), and $Traj_{c}$ denotes the trajectory of ADS with the cyber configuration $c$. This objective function is necessary, otherwise one can just set the ADS's speed to zero and the accident can be suppressed.
\begin{equation}\label{eq:h1}
\begin{split}
    h_1(c) = - Sim(Traj_{c_0}, Traj_c)
\end{split}
\end{equation}
\indent $h_2(\cdot)$ has the same purpose as $f_3(\cdot)$ in Eq.~\ref{eq:f3}. We do not repeat details here due to space concerns.
\begin{equation}
\begin{split}
    h_2(c) = \#(c-c_0) + \sum_{k \in Enum} L_1 (k_0, k) + \sum_{k \notin Enum} L_2 (k_0, k)
\end{split}
\end{equation}\label{eq:h2}

%% file: figtex/msg-reverse-graph.tex
\begin{figure*}
    \begin{minipage}{.65\linewidth}
        \centering
        \includegraphics[width=\linewidth]{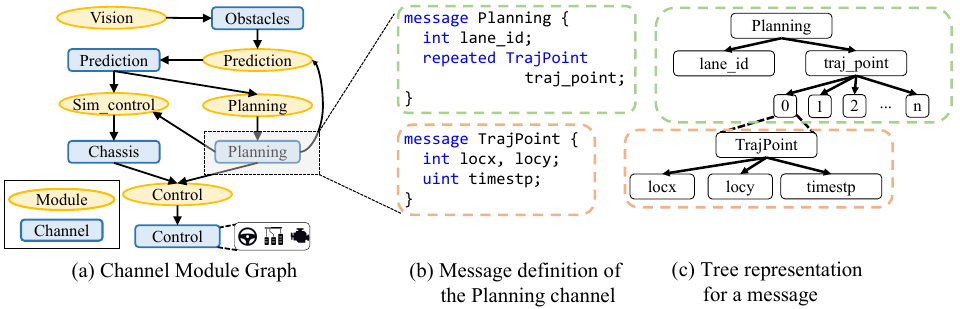}
        \caption{How \toolname represents one execution. In (a), yellow ovals denote modules and blue boxes denote channels.
        }
        \label{fig:running-ex}
    \end{minipage}
    ~\hfill~
    \begin{minipage}{.26\linewidth}
        \centering
        \includegraphics[width=\linewidth]{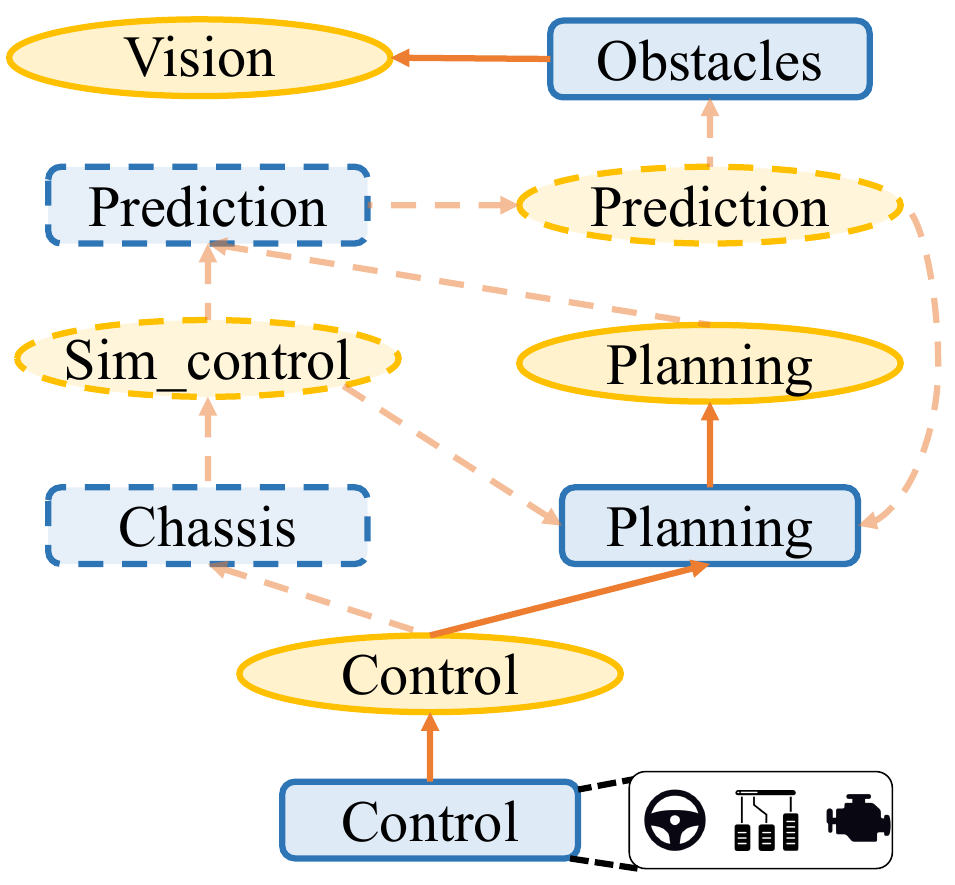}
        \caption{Reversed CMG in Alg.~\ref{alg:init-module}. 
        }
        \label{fig:causality-graph}
    \end{minipage}
    \vspace{-10pt}
\end{figure*}

%% file: figtex/init-module.tex
\begin{algorithm}[t]
\footnotesize
\caption{Pinpoint initial deviating module}\label{alg:init-module}
\Func{
    \func{Find\_Init\_Module} (G, channel2mdr) 
}{ 
    \tcp{\textrm{{\it channel2mdr}}: \textnormal{map from channels to \metric{MDR} series}}
    \tcp{$t^*$: \textnormal{map from node to its earliest changing point}}
    $t^*=\emptyset$ 
    \\
    $\textrm{{\it channel2chng\_pt}}=\func{PELT}(\textrm{{\it channel2mdr}})$\\
    \For{node $\in$ G.nodes}{
        \If{node is Channel}{
            $t^*$[node] = \textrm{{\it channel2chng\_pt}}[node][0]
        }
        \If{node is Module}{
            $t^*$[node] = $\textrm{min}_{(node, ch) \in G.edges}$ $t^*$[ch]
        }
    }

    \For{(u,v) $\in$ G.edges}{
        \If{$t^*(u) \notin [t^*(v)-\delta, t^*(v)]$}{
            G.\func{remove\_edge}($u,v$)\\
        }
    }
    $ctrlnode$ = \var{Control} channel node\\
    $D$ = \{ $m$ | $m$ is Module and reachable from $ctrlnode$ in reversed $G$ \} \\
    \Return{ $\textrm{argmin}_{m \in D}$ $t^*(m)$ }\\
}
    
\end{algorithm}

%% file: 6-evaluation.tex
\input{tables/case_list}

\section{Evaluation}

We introduce our evaluation setup (Section~\ref{sec:eval-setup}) and present results by answering the following research questions (RQs):\\
{\bf RQ1:} How effective is \toolname on finding triggering entities and misconfigurations? (Section~\ref{sec:eval-effectiveness})\\
{\bf RQ2:} How effective is the proposed metric \metric{MDR}? (Section~\ref{sec:eval-mdr}) \\
{\bf RQ3:} How efficient is \toolname and each phase? (Section~\ref{sec:eval-efficiency}) \\
{\bf RQ4:} How does AD root cause analysis help the accident investigation? (Case studies in Section~\ref{sec:eval-case-studies})

\subsection{Setup}\label{sec:eval-setup}

\noindent \textbf{Implementation.} \toolname includes several components: 
(1) The replay engine that can replay actor trajectories inside simulator;
(2) The physical mutator that can change the behaviors of actor entities and map entities; 
(3) The graph extractor that dynamically extracts CMG.  
(4) The execution differentiation algorithm that identifies the initial deviating module; 
(5) The cyber mutator that pinpoints the misconfiguration of the subject ADS. We run experiments on Ubuntu 20.04, with 96 GB RAM and Nvidia GPU 3090.

\noindent \textbf{Subject Systems.} 
We evaluate \toolname on two open-source ADSs, Baidu Apollo~\cite{baiduapollo} and LAV~\cite{chen2022lav}, both designed in a modularized paradigm. 
Apollo is representative for industry-grade ADS, as Baidu has obtained permits to operate fully autonomous taxis without any human assistance in China since August 2022. 
LAV is a research-oriented ADS. We use it to show that \toolname can be generalized to other modularized ADS, even if it is not based on ROS-like communication frameworks. Specifically, we manually isolate different modules by inspecting the code and instrument the data transmitted between modules. Such data approximates the channel messages in an ROS-like framework. The manual work is a one-time effort and affordable (2 man-hours). For simulation, we use the simulator supported by the corresponding ADS, namely LGSVL~\cite{lgsvl-sim} for Apollo and CARLA~\cite{carla-sim} for LAV.

\noindent \textbf{Logging and Replay.}
During ADS's execution, it is crucial that \toolname records the
information of other actor entities.
During the post-accident replay, 
\toolname needs to extract the locations and rotations of each actor entity from the recorded messages. 
Since these values are usually in the AD system's coordinates, \toolname transforms them into the simulator's world coordinates, using the transformation matrices of respective simulators.

\noindent \textbf{Accident Cases.}  
We investigate 184 accident cases from recent literature~\cite{tang2021ase,huai2023doppeltest,planfuzz,drivefuzz,tang2021iv,tang2021icra}. 
These cases had been confirmed and hence can serve as the ground truth. We consider our tool correctly identifies a root cause if it locates the same misconfiguration.
We categorize them via 3 perspectives: {\ding{182}} whether it can be exploited by leveraging external scenario
, {\ding{183}} whether the accident consequence is severe (i.e., collision or emergency braking) or just efficiency degradation (e.g., taking longer routes), {\ding{184}} what the driving scenario is at accident moment. Detailed statistics are shown in Table~\ref{tab:case-stats} (in Supplementary Material of~\cite{rocas-github-repo}).
In order to eliminate non-safety-related incidents, we filter out cases that are not exploitable and solely related to performance issues. After the filtering process, as shown in Table~\ref{tab:case-select} (in Supplementary Material of~\cite{rocas-github-repo}), 144 cases are left, which we further categorize into 12 types based on similarities in ADS's behaviors, accident consequences, and driving scenarios. Table~\ref{tab:caselist} shows a comprehensive overview of these categorized cases.

\input{tables/replay_n_caseselect}

\input{tables/case_root_cause}

\subsection{Effectiveness of Root Cause Analysis}\label{sec:eval-effectiveness}

\noindent \textbf{Accident Root Cause Analysis.}
Table~\ref{tab:case-root-cause} consists of several columns that provide essential information regarding the accidents and the corresponding investigation results from each phase. There are three main categories based on the identified deviating module. (1) Prediction related accidents. In A1, the ADS incorrectly predicts a moving car as static, and in A10 the ADS predicts the wrong trajectory of a right-side cyclist. 
(2) Planning related accidents. A2, A6, and A11 are different scenarios where the ADS has the wrong decision on whether other entities have overlap with the ADS's planned trajectory. A3, A4, and A5 are different scenarios where the ADS has the wrong decision on whether it should overtake or yield to other entities.
(3) Perception related accidents. In A9, the ADS fluctuates the braking probability when seeing a red truck ahead. In A12, due to the special road shape, the ADS recognizes the traffic light from the wrong direction.  We further illustrate the accident scenarios for the cases in Table~\ref{tab:caselist} in Figure~\ref{fig:case-imgs} in the Supplementary Material of~\cite{rocas-github-repo} (A7 in Table~\ref{tab:caselist} has been illustrated in Figure~\ref{fig:moti}). 

\noindent \textbf{Regression Test.} We conduct regression tests to ensure that our mutated configurations do not introduce new hazardous behaviors. We constructed a regression dataset with 200 cases by adding small perturbations to the original executions. In our evaluation, all modified configurations achieved higher mission success rates than the original configurations.

\noindent \textbf{Failure Cases.} 
\toolname has certain limitations when dealing with two specific accident types, namely Type A7 and A8. Type A7 involves collisions with other vehicles approaching the stationary ADS from lateral or rear directions. In such cases, \toolname can identify the triggering entity during the physical mutation phase but is unable to modify a configuration value in the cyber mutation phase to prevent the accident. This limitation arises because most ADSs do not incorporate features to prevent collisions from lateral or rear directions, as such collisions are generally considered inevitable and not within the typical expectations for ADSs to handle.
As for Type A8, it pertains to collisions with road-side curbstones or vegetation that occur when the ADS follows abnormal trajectories. Upon manual investigation, we found that these collisions occur when there is no feasible path from the current AD's location to the routing destination. As a result, the ADS randomly deviates from its original trajectories. This situation may arise when the AD vehicle is spawned at a junction or roads with opposite directions, or when the AD vehicle is blocked at certain corners where the ADS lacks a back-up feature to navigate out of the situation.
In summary, the collision incidents in Type A7 and A8 are primarily linked to the lack of feature implementation, such as the absence of measures to avoid lateral collisions or handle infeasible routing, rather than being caused by misconfigurations.

\subsection{Effectiveness of \metric{MDR}} \label{sec:eval-mdr}

 \indent Table~\ref{tab:eval-scope} shows \toolname in average reduces search space of configurations by 85.8\%.
Table~\ref{tab:eval-scope} provides a detailed breakdown of the number of configurable parameters in Apollo's four main modules, including the perception, prediction, planning, and control module (Rows 2-5). The sum of configurable parameters across all modules is also presented in Row 6. 
We can see that, due to the complexity of the Apollo system, there are a significant number of configurable parameters, which can lead to misconfigurations and subsequent failures. However, \toolname has proven to be an effective tool for narrowing down the search space and identifying these misconfigurations. 
By leveraging its execution diff algorithm, \toolname successfully reduces the misconfiguration search space to below 20\% of the original whole search space. 
This significant reduction in search space contributes to the runtime efficiency of  Phase-III (cyber mutation).
We also include a concrete example in Supplementary Material of~\cite{rocas-github-repo} Section~\ref{sec:app:how-narrow} to illustrate how \toolname narrows down the search space.

\subsection{Efficiency of Runtime} \label{sec:eval-efficiency}

\begin{figure}
    \centering
    \includegraphics[width=0.95\linewidth]{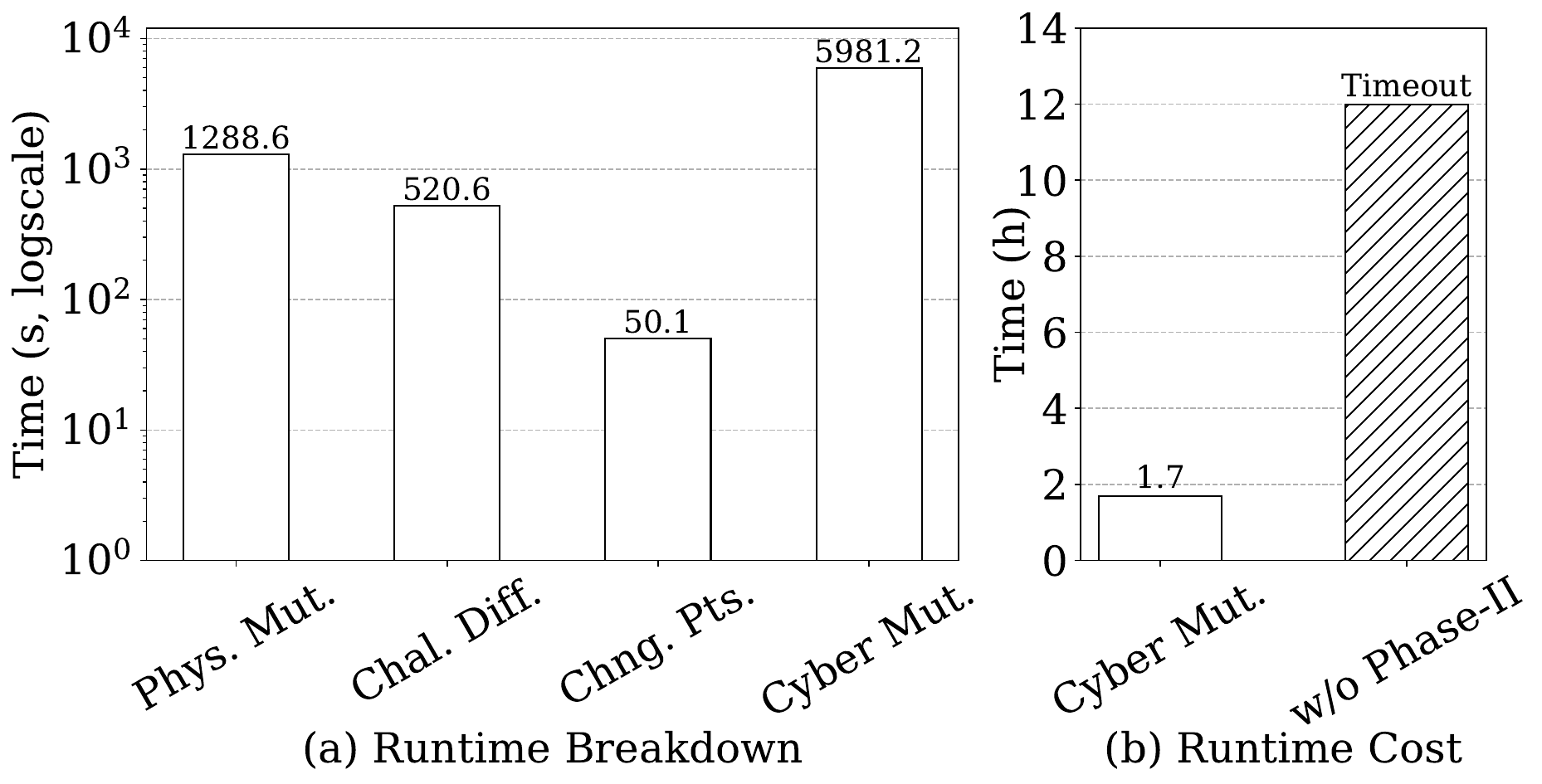}
    \caption{Runtime. (a) shows runtime of each phase. (b) shows the benefit of search scope reduction.}
    \label{fig:runtime}
\end{figure}

Because the length of an execution can impact the investigation runtime (for instance, a longer execution can result in a larger mutation search space and more execution records to analyze), we trim executions to the last 30 seconds before the occurrence of accidents. By doing so, we can limit the amount of data we need to analyze while still capturing the critical period leading up to the accidents. The average storage size for a 30-second execution is around 30 MB.
The experimental results show that Phase-III (i.e., the cyber mutation) represents 76.3\% of the total time. It demonstrates that the scope reduction of Phase-II is necessary to make the root cause analysis affordable. On the other hand, a baseline method without scope reduction cannot find any misconfiguration within a time budget of 12 hours. Details are discussed below.\\
\noindent \textbf{Time Breakdown.} Figure~\ref{fig:runtime}a displays the average runtime cost for each step during an accident investigation. Phase-I (physical mutation) takes up 16.4\% of the total runtime, while Phase-II (including channel message differential analysis and change point detection) accounts for 7.3\%. The most time-consuming part is Phase-III (cyber mutation) which takes up 76.3\% of the total time.

The reason why Phase-I is much faster than Phase-III is that it is easier to identify a mutation that can prevent accidents when modifying physical entities (e.g., changing the location or speed of other vehicles). However, in Phase-III, \toolname must identify the responsible mis-configuration among many possible parameters to avoid the accident. Therefore, the scope reduction of Phase-II is necessary to make the root cause analysis affordable.

\noindent \textbf{Compare with Baseline.} Since there is no existing work that can be directly applied on the ADS root cause analysis task, we compare \toolname with a naive brute force method that searches among all ADS modules.
Figure~\ref{fig:runtime}b shows the average runtime comparison for cyber mutation, with and without the scope reduction. The naive brute force method cannot find a misconfiguration within a reasonable budget (12 hours).

\subsection{Case Studies} \label{sec:eval-case-studies}

We present the details of two cases (A4 and A6 in Table~\ref{tab:caselist}) to demonstrate the benefits of \toolname 
and how the found misconfiguration can shed light on the mystery behind the accidents. 
Case Study I and II are in Supplementary Material of~\cite{rocas-github-repo} Section~\ref{sec:app:case-study}.

%% file: tables/case_list.tex
\begin{table*}[!t]
    \centering
    \small
    \setlength{\tabcolsep}{1.5pt}
    \caption{
    Accident Types Summary. Fig.~\ref{fig:case-imgs} illustrates scenarios. Link~\cite{case-youtube-link} shows accident videos.
    }
    \label{tab:caselist}
    \begin{threeparttable}
    \begin{tabular}{cccclcc}
        \toprule
            {\bf Type} & 
            {\bf Src.${\dagger}$} & 
            {\bf \#Inst.} & 
            {\bf SUT} & 
            {\bf Description} &
            {\bf Conseq.} & 
            {\bf Scenario} 
            \\
        \midrule
        A1 & CA/DF/DT & 37 & Apollo 7.0 & {Incorrectly takes a slowly moving car as a static object} & 
        Collision & Intersection     \\
        \midrule
        A2 & PF & 3 & Apollo 7.0 & {Blocked by 2 stopped vehicles ahead although there is enough space in between} & 
        EB & Following      \\
        \midrule
        A3 & CA/AT & 3 & Apollo 7.0 & {Turns left at junction and collides with a truck from opposite direction when trying to overtake it} & 
        Collision & Intersection          \\
        \midrule
        A4 & AT/DF/DT & 43 & Apollo 7.0 & {Over-aggressively overtakes a left-turning vehicle} &
        Collision & Merging  \\
        \midrule
        A5 & CA/AT & 36 & Apollo 7.0 & {Oscillates between overtake and yield and cannot stop in time when finally deciding to yield} & 
        Collision & Intersection        \\
        \midrule
        A6 & DF & 2 & Apollo 7.0 & {Fails to avoid collision with a large truck drifting on wet ground} &
        Collision & Intersection        \\
        \midrule
        A7 & DF & 4 & Apollo 7.0 & {Collides with other vehicles from lateral or rear direction} &
        Collision & Merging   \\
        \midrule
        A8 & DF & 3 & LAV & {Collides with road-side curbstone or vegetation} &
        Collision & Following   \\
        \midrule
        A9 & DF/DT & 5 & LAV        & {Takes a red truck as red light and suddenly stops in roundabout} &
        EB & Intersection     \\
        \midrule
        A10 & DF & 2 & LAV        & {Recognizes the wrong traffic light due to special road shape and suddenly stops on the road} & 
        EB & Turning      \\
        \midrule
        A11 & DF & 2 & LAV       & {Turns right and fails to avoid collision with a wide truck} & 
        Collision & Merging  \\
        \midrule
        A12 & DF/PF & 4 & LAV        & {Predicts the wrong trajectories of a cyclist on the side of AD vehicle} & 
        EB & Turning  \\
        \bottomrule
    \end{tabular}
    \begin{tablenotes}[para,flushleft]
        Src: Accident Source,
       \#Inst.: Instance Number,
       SUT: System Under Test,
       Conseq.: Consequences,
       EB: Emergency Braking.
    \end{tablenotes}
    \begin{tablenotes}[para, flushleft]
       $\dagger$ Accident Sources. 
       CA: CAT~\cite{tang2021iv}, 
       AT: ATLAS~\cite{tang2021ase}, 
       DF: DriveFuzz~\cite{drivefuzz}, 
       DT: DoppelTest~\cite{huai2023doppeltest},
       PF: PlanFuzz~\cite{planfuzz}.
    \end{tablenotes}
    \end{threeparttable}
    \vspace{-10pt}
\end{table*}

%% file: tables/replay_n_caseselect.tex
\begin{table}[]
    \centering
    \small
    \setlength{\tabcolsep}{2pt}
    \renewcommand{\arraystretch}{1}
    \captionof{table}{Scope reduction of configurations in Baidu Apollo.}\label{tab:eval-scope}
    \begin{tabular}{lcc}
         \toprule
         \textbf{Main Modules} & \textbf{\# Config.} & \textbf{Proportion} \\
         \midrule
         Perception & 171 & 15.35\% \\
         Prediction & 183 & 16.53\% \\
         Planning   & 222 & 19.93\% \\
         Control    &  55 &  4.94\% \\
         \midrule
         All Modules       & 1114 & 100\% \\
         \bottomrule
    \end{tabular}
\end{table}

%% file: tables/case_root_cause.tex
\begin{table*}[ht]
    \centering
    \small
    \setlength{\tabcolsep}{2.5pt}
    \caption{
    Root Cause Analysis Results. 
    Physical Mutation 4 columns, display Phase-I output (i.e., the mutated value of entity properties v.s. original values).
    Deviating Module column shows Phase-II output.
    Cyber Mutation 3 columns, display Phase-III output (i.e., the mutated value of configurations v.s. original values). 
    }
    \label{tab:case-root-cause}
    \begin{threeparttable}
    \begin{tabular}{cccccccccc}
        \toprule
            \multirow{2.5}{*}{\bf Type} & 
            \multirow{2.5}{*}{\bf System} & 
            \multicolumn{4}{c}{\bf Physical Mutation} & 
            \multirow{2.5}{*}{\makecell[c]{\bf Deviating\\ \bf Module}}  & 
            \multicolumn{3}{c}{\bf Cyber Mutation }   \\
        \cmidrule(lr){3-6}\cmidrule(lr){8-10}
          & & Triggering Entity &   Property & Ori. Val. &  Mut. Val. &  &   Misconfiguration &   Ori. Val. &    Mut. Val. \\
        \midrule
        A1 & Apollo 7.0 & Collided Sedan & Speed & 0.98 & 1.2 & Prediction & \var{still\_obstacle\_speed\_threshold} & 0.99 & 0.50       \\
        \midrule
        A2 & Apollo 7.0 & Right-front Sedan & Location.x & 14.42 & 14.37 & Planning & \var{obstacle\_lat\_buffer}  & 0.60 & 0.20       \\
        \midrule
        A3 & Apollo 7.0 & Collided Truck          & \makecell[c]{Rotation.yaw\\Location.y} & \makecell[c]{275\\-27.1} & \makecell[c]{285\\-27.3} & Planning & \var{kMinOvertakeDistance} & 10.0 & 40.0    \\
        \midrule
        A4 & Apollo 7.0 & Collided Sedan & Speed & 6.0 & 4.0 & Planning & \var{yield\_distance} & 5.0 & 2.0 \\
        \midrule
        A5 & Apollo 7.0 & Collided Sedan & Location.x & -234.5 & -235.2 & Planning & \var{kOvertakeTimeBuffer} & 3.0 & 2.0        \\
        \midrule
        A6 & Apollo 7.0 & Collided Truck & Rotation.yaw & 310 & 300 & Planning & \var{kADCSafetyLBuffer} & 0.1 & 1.0        \\
        \midrule
        A7 & Apollo 7.0 & Collided Sedan & Location.x & 13.5 & 13.0 & Perception & - & - & -        \\
        \midrule
        A8 & LAV & Collided Building & - & Enabled & Disabled & Perception & - & - & -        \\
        \midrule
        A9 & LAV & Front-side truck & --  & Enabled & Disabled & Perception & \var{brake\_threshold} & 0.10 & 0.20       \\
        \midrule
        A10 & LAV & \makecell[c]{Left-side\\Traffic Light} & Policy & Red & Green & Perception & \var{brake\_threshold}  & 0.10 & 0.90       \\
        \midrule
        A11 & LAV & Left-front Truck & Rotation.yaw & 32 & 25 & Planning & \var{dist\_threshold\_moving} & 2.50 & 3.50  \\
        \midrule
        A12 & LAV & Right-side cyclist & Speed & 1.0 & 0.0 & Prediction & \var{dist\_threshold\_static} & 1.00 & 0.50  \\
        \bottomrule
    \end{tabular}
    \end{threeparttable}
    \vspace{-10pt}
\end{table*}

%% file: 7-discussion.tex
\section{Discussions}\label{sec:discuss}
\noindent \textbf{Practicality.} We use simulated accidents instead of real-world accidents for the following reasons. Firstly, accidents data of real-world deployed ADS are typically considered proprietary by companies and not publicly available. Secondly, conducting real-world experiments on accident investigation is often prohibitively expensive. Additionally, recent advances in high-fidelity simulators~\cite{carla-sim, lgsvl-sim, waymo-sim} enable simulations to closely reflect real-world conditions.

\noindent \textbf{Limitations.} 
\toolname{} is limited to modularized ADS and cannot be applied to the end-to-end (e.g., reinforcement learning) ADS. In the latter case, \toolname can only identify the triggering entity, but it cannot pinpoint the specific module or misconfiguration. 

\noindent \textbf{Validation.} \toolname outputs a misconfiguration, but it still requires human to inspect misconfiguration usages to finally understand the rationale behind accidents. Also, \toolname can help {\it finetune}, but not {\it fix} misconfigurations of ADS. Fixing configuration issues is complimentary to \toolname and still an open challenging problems in SE community~\cite{han2016studyperform, han2018perflearner, ha2019deepperf, he2020cpdetector, krishna2020cadet, Velez2022performconfig, de2022different, valle2023cpsmisconfigrepair}. We leave it as future work. \looseness=-1

%% file: 8-related.tex
\section{Related Work}

\noindent \textbf{CPS Fuzzing.} Inspired by traditional fuzzing methods~\cite{afl, bohme2019greyfuzz,xie2019CGF, wen23multilangfuzz}, recently many CPS fuzzing techniques have been proposed to search for vulnerabilities in CPS, including drone systems~\cite{kim2019rvfuzzer,kim2021pgfuzz,han2022detectspecbugs}, ADS~\cite{garcia2020avbugs,Zampetti2022cpsbug,zhong2021scenariosurvey,khan2023wip,humeniuk2022search,pang2022mdpfuzz,zhong2022detectfusion,planfuzz,drivefuzz,song2023acero,cheng2024fusion,cheng2024badpart, wang2024dance}, ROS~\cite{xie2022rozz,kim2022robofuzz}, DNN controllers~\cite{jung2022swarmflawfinder}, 
smart homes~\cite{mandal2023helion}, satellites~\cite{lee2023fuzzcpsmutation, menghi2021trace} 
and anti-terrorism system~\cite{Cascavilla2020Counterterrorism}. 
Our post-accident analysis is complementary to the fuzzing methods mentioned above. \looseness=-1

\noindent \textbf{Root Cause Analysis.} Root cause analysis is critical for debugging program failures.
Numerous techniques have been proposed, such as 
log-based causality analysis~\cite{goel2005taser,krishnan2010trail,kim2010intrusion,king2003backtracking,lee2013high}. Some use program instrumentation to generate execution logs~\cite{ohmann2017lightweight,yuan2010sherlog}, while others focus on recording OS events during runtime and performing offline analysis.
Techniques like taint analysis~\cite{bosman2011minemu,clause2007dytan,kang2011dta++} and record-and-replay provenance analysis~\cite{walkup2020forensic,chandra2011intrusion,goel2005taser,guo2008r2,srinivasan2004flashback,narayanasamy2006bugnet,kim2010intrusion,geels2007friday} are also widely used for accident investigation. However, these approaches induce heavy overhead, and are not suitable for resource-constrained CPS. 

\noindent \textbf{Fault localization.} Fault localization techniques~\cite{wong2016survey} vary widely, including slice-based~\cite{weiser1984program}, spectrum-based~\cite{harrold2000empirical}, statistics-based~\cite{zheng2006statistical}, states-based~\cite{zeller2002state}, learning-based~\cite{brun2004finding}, etc. Our method differs from existing fault localization methods in two aspects: (1) Fault localization methods typically pinpoint suspicious code element (e.g., statements, predicates, functions or files), while our method pinpoints the misconfiguration. (2) Fault localization methods focus on program bugs, while our focus is on misconfiguration, which does not necessarily mean that the code logic is buggy.

\noindent \textbf{CPS Root Cause Analysis.} Root cause analysis of CPS is a less investigated area. There are existing efforts focusing on drone systems~\cite{clark2017drop,jain2017drone,kim2020mayday,choi2022rvplayer}. They either fail to capture complex behaviors across multiple modules in ADS~\cite{clark2017drop,jain2017drone,choi2022rvplayer} or requires heavy-weight program instrumentation
and trace collection~\cite{kim2020mayday}. 
A recent work CARE~\cite{hossen2023care} focuses on mission failures in robots, it cannot reason environments with moving objects. On the other hand, Swarmbug~\cite{jung2021swarmbug} focuses on problems in the swarm coordination, not considering internal logic of individual agents.
Compared with the above techniques, our work is the first to target at AD scenarios that can handle both more complex physical environments and internal program configurations.

%% file: 9-conclusion.tex
\section{Conclusion}

In this paper, we first formally define the problem of ADS root cause analysis and present \toolname{}, a novel method for ADS root cause analysis.
Our technique leverages both physical and cyber mutation that can
precisely identify the accident-trigger entity and pinpoint the misconfiguration responsible for an accident.
We demonstrate the effectiveness and efficiency of \toolname{} through the evaluation of 12 types of ADS accidents, 144 accidents in total.

\section*{Acknowledgement}
We thank the anonymous reviewers for their valuable comments and suggestions. 
We are grateful to the Center for AI Safety for providing computational resources. This work was funded in part by the National Science Foundation (NSF) Awards SHF-1901242, SHF-1910300, IIS-2416835, DARPA VSPELLS - HR001120S0058, IARPA TrojAI W911NF-19-S0012, ONR N000141712045, N000141410468 and N000141712947. Any opinions, findings and conclusions or recommendations expressed in this material are those of the authors and do not necessarily reflect the views of the sponsors.

%% file: 10-appendix.tex
\newpage

\appendix
\section*{Appendix}

\setcounter{section}{0}
\renewcommand*{\theHsection}{appendix.\the\value{section}}

\section{How \toolname Reduces Search Space}
\label{sec:app:how-narrow}
\begin{figure*}
    \begin{minipage}{.4\linewidth}
        \centering
        \includegraphics[width=\linewidth]{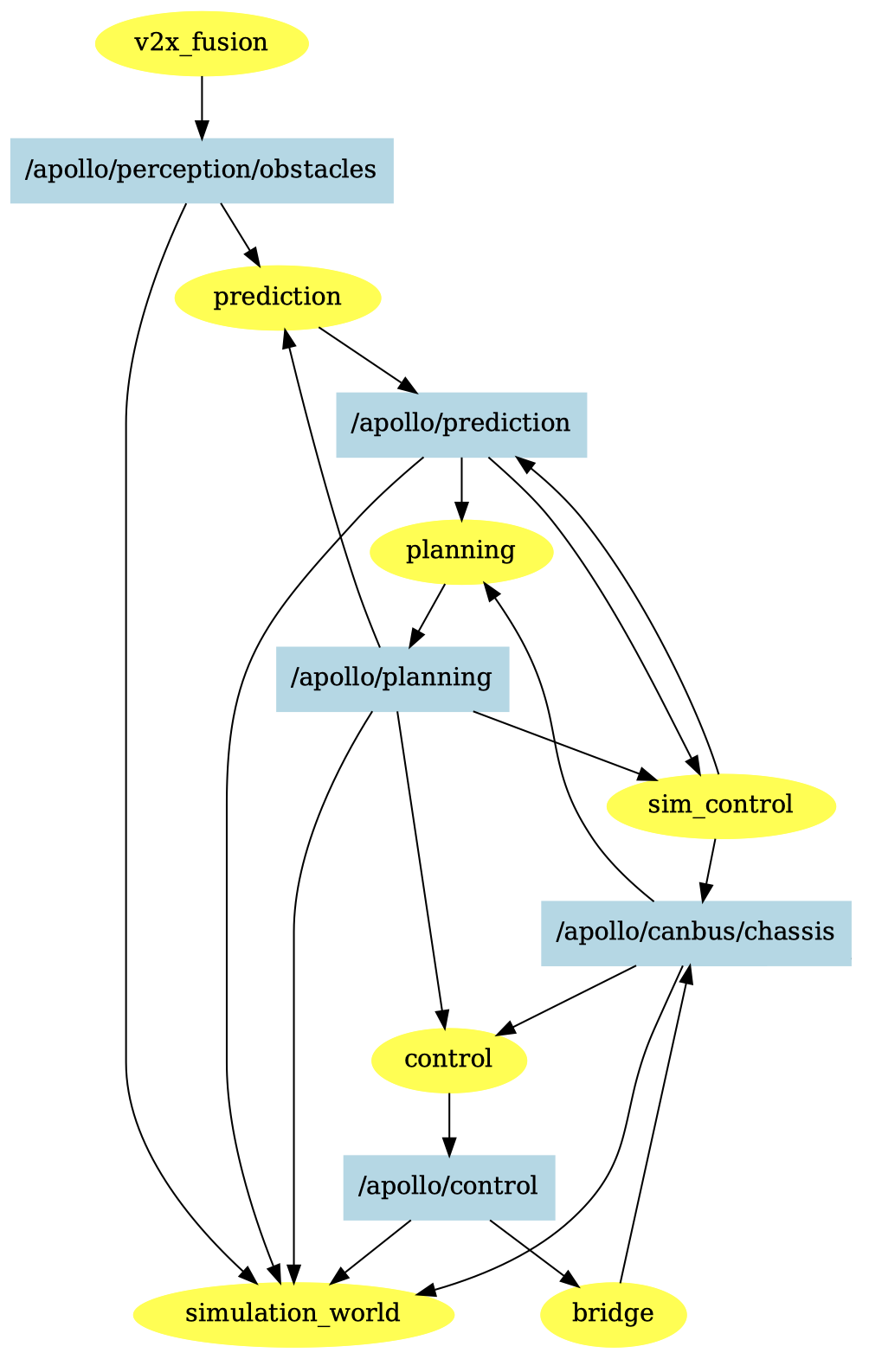}
        \caption{Extracted (simplified) CMG from CyberRT.}
        \label{fig:eval-graph-node}
    \end{minipage}
    \hspace{.05\linewidth}
    \begin{minipage}{.5\linewidth}
        \centering
        \begin{subfigure}{\linewidth}
        \centering
        \includegraphics[width=0.7\linewidth]{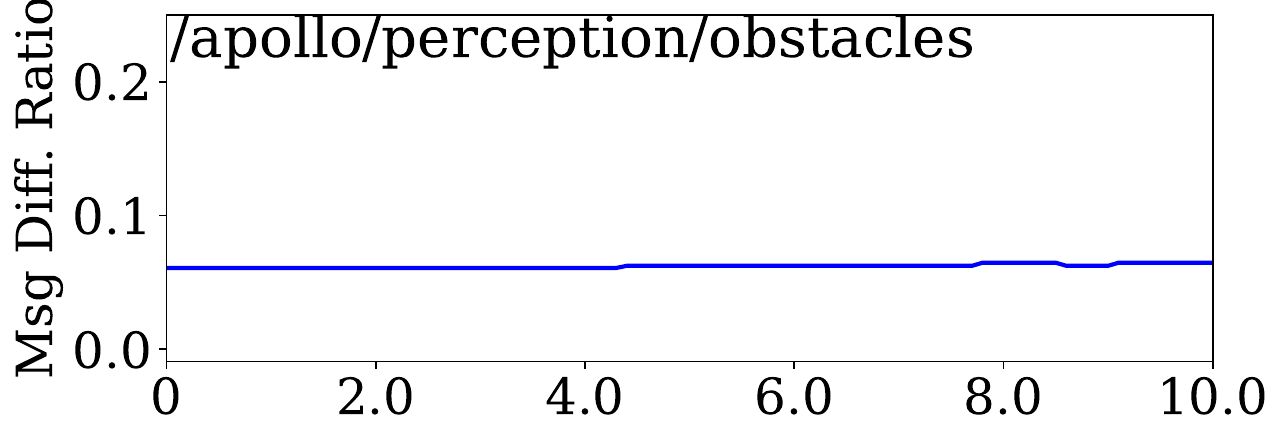}
        \end{subfigure}
        ~
        \begin{subfigure}{\linewidth}
        \centering
        \includegraphics[width=0.7\linewidth]{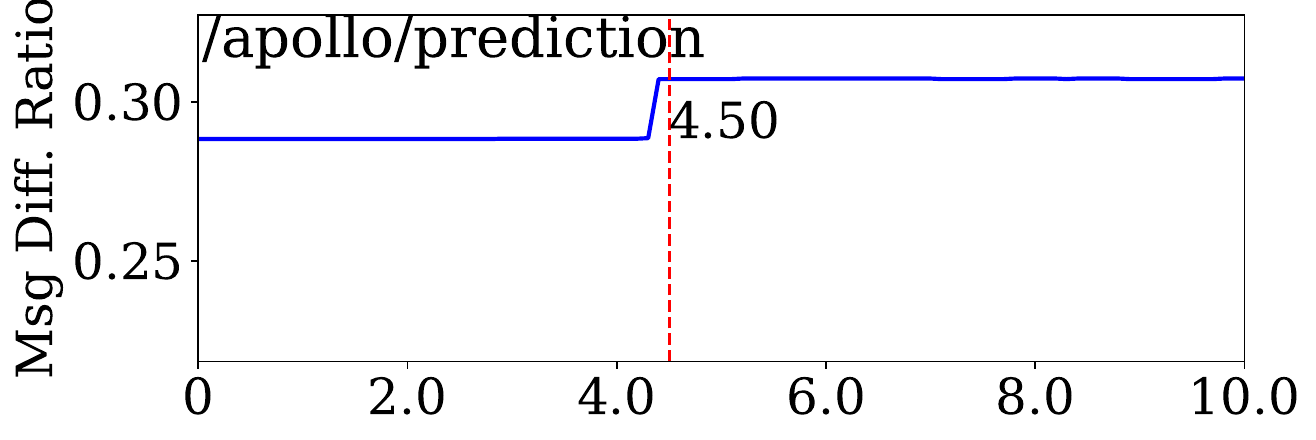}
        \end{subfigure}
        ~
        \begin{subfigure}{\linewidth}
        \centering
        \includegraphics[width=0.7\linewidth]{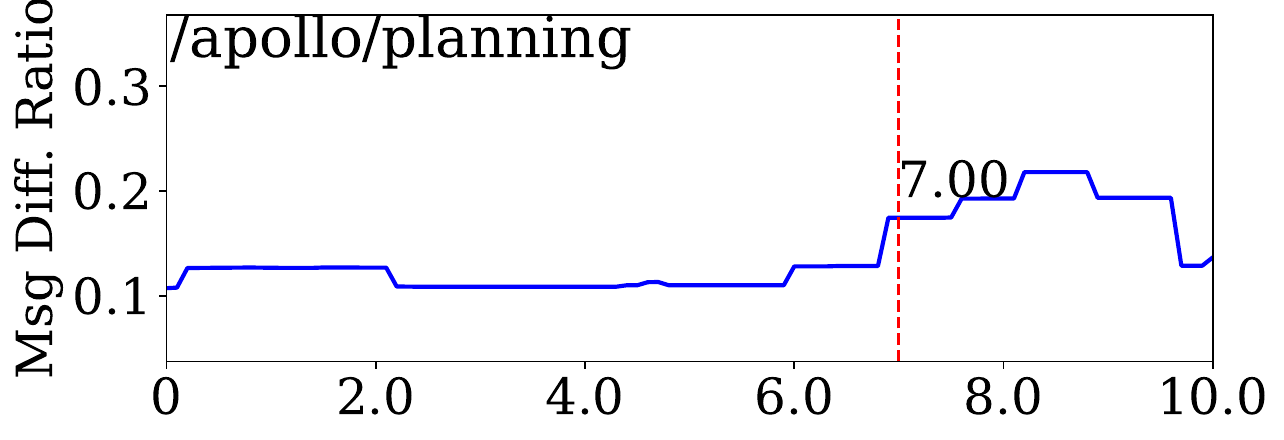}
        \end{subfigure}
        ~
        \begin{subfigure}{\linewidth}
        \centering
        \includegraphics[width=0.7\linewidth]{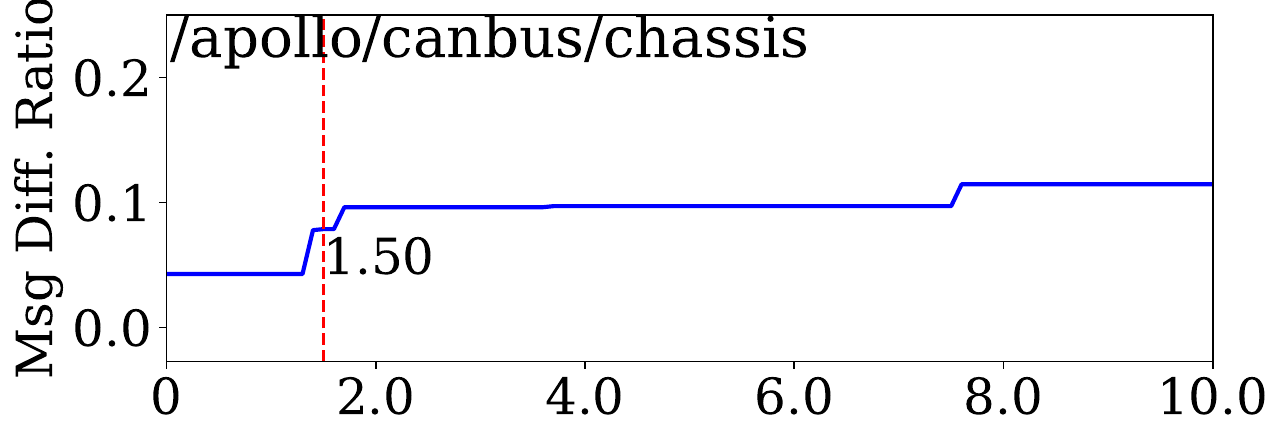}
        \end{subfigure}
        ~
        \begin{subfigure}{\linewidth}
        \centering
        \includegraphics[width=0.7\linewidth]{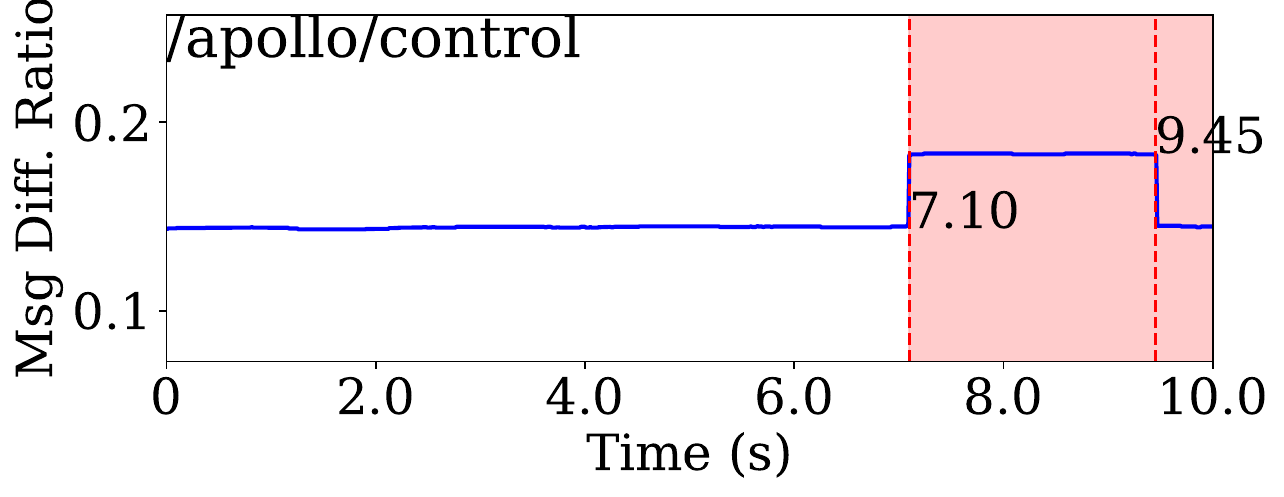}
        \end{subfigure}
        \caption{
            \metric{MDR} series for CMG channels in Fig.~\ref{fig:eval-graph-node}. Red dot lines denote \metric{MDR} change points. The accident happens at 7.1s, shown as red region in \textit{control} channel subfigure.
        }
        \label{fig:eval-channel-mdr}
    \end{minipage}
    \vspace{5pt}
\end{figure*}

As we mentioned in Section~\ref{sec:design-id-module}, due to the natural noise of the physical environment, there always exist some differences when conducting the differential analysis (in Section~\ref{sec:diff-exec}). Therefore it is necessary and crucial to locate the accident-inducing difference. 
Figure~\ref{fig:eval-channel-mdr} shows the \metric{MDR} series of different channels (for accident A1), corresponding to the CMG in Figure~\ref{fig:eval-graph-node}. Red dot lines denote the change points. This accident happens at 7.1s, marked as the red region in the \var{control} channel subfigure.\\
\indent It is worth noting that different channels have different \textit{default} \metric{MDR} values. For example, the \var{planning} channel has a default \metric{MDR} of 12.84, while \var{prediction} channel has a default \metric{MDR} of 28.86, as shown in Table~\ref{tab:eval-channel-cgpt}.
When running Algorithm~\ref{alg:init-module} (with $\delta=3$) to find the deviating module, we can obtain a path in reversed CMG, namely \var{control} $\rightarrow$ \var{planning} $\rightarrow$ \var{prediction}. Therefore module \var{prediction} is correctly identified. Although channel \var{chassis} also has a change point at 1.5s, it is not reachable from channel \var{control} because the edge from channel \var{chassis} to module \var{control} has been removed at Algorithm~\ref{alg:init-module} Line 11.\\

\section{Case Studies}~\label{sec:app:case-study}
\input{figtex/case-study1}

\input{figtex/case-study2}

\begin{table}[t]
        \centering %
    \setlength{\tabcolsep}{2pt}
    \small
    \caption{
        Selecting Criteria.
        From Table~\ref{tab:case-stats} we select exploitable and severe cases (i.e., collision or emergency braking). 
    }\label{tab:case-select}
        \vspace{2pt}
    \begin{tabular}{r|cc|c}
        \toprule
          &  Inefficient & Severe & All \\
          \midrule
         Exploitable     & 13 & {\bf 144} & 157 \\
         Inexploitable    & 12 & 15 & 27 \\
         \midrule
         All & 25 & 159 & 184 \\
         \bottomrule
    \end{tabular}
    \hspace{.01\linewidth}
    \vspace{5pt}
\end{table}

\begin{table}[t]
        \centering 
        \setlength{\tabcolsep}{1.5pt}
        \small
        \captionof{table}{Change time point $t^*$ for each channel and the \metric{MDR} value before and after \metric{MDR}.} 
        \label{tab:eval-channel-cgpt}
        \begin{tabular}{lcccc}
        \toprule
        \multirow{2.5}{*}{\textbf{Channel Name}} & \multirow{2.5}{*}{{\bf $t*$}}  & & {\bf \metric{MDR}} \\
        \cmidrule(lr){3-5}
             & & \textbf{Before} & \textbf{After} & \textbf{$\Delta$ Ratio}  \\
        \midrule
             /apollo/perception/obstacles   & - &  -  & - & -  \\
             /apollo/prediction             & 4.50 & 28.86  & 30.72 & +1.86  \\
             /apollo/planning               & 7.00 & 12.84  & 17.44 & +4.60  \\
             /apollo/canbus/chassis         & 1.50 & 4.28   & 9.62  & +5.35  \\
             /apollo/control                & 7.10 & 14.52  & 18.29 & +3.78  \\
        \bottomrule
        \end{tabular}
        \vspace{2pt}
\end{table}

\input{tables/case_stats}

\begin{figure*}[!h]
    \centering
    \includegraphics[width=0.95\linewidth]{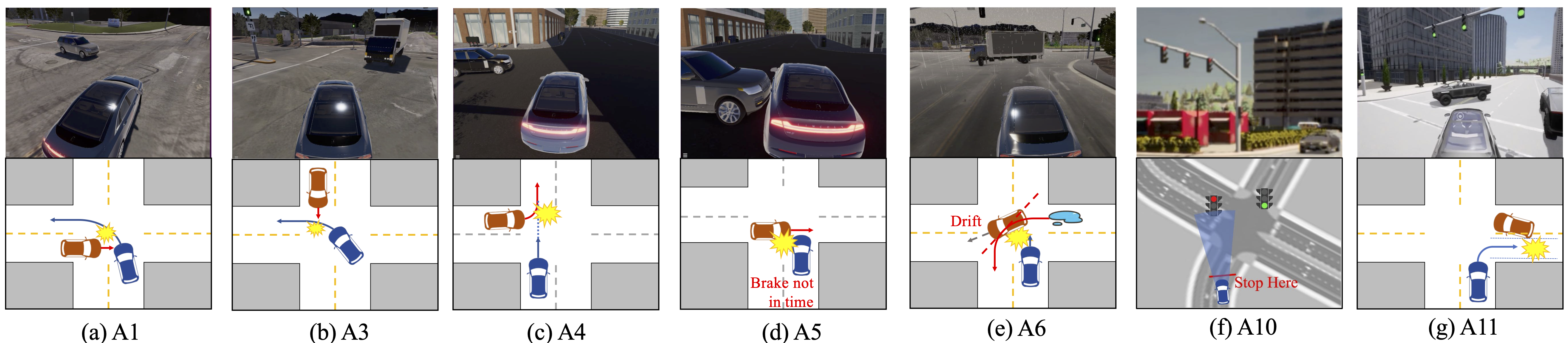}
    \caption{Illustrations for accident scenarios in Table~\ref{tab:caselist}. Top images show simulator screenshots just before or at the accident moments. Blue vehicles are the ADSs. Solid lines show vehicles' trajectories. Vehicles without lines mean their speeds are 0.
    }
    \label{fig:case-imgs}
\end{figure*}

\newpage

%% file: figtex/case-study1.tex
\begin{figure*}[h!]
    \centering
    \begin{subfigure}[t]{0.31\linewidth}
        \centering
        \includegraphics[width=\linewidth]{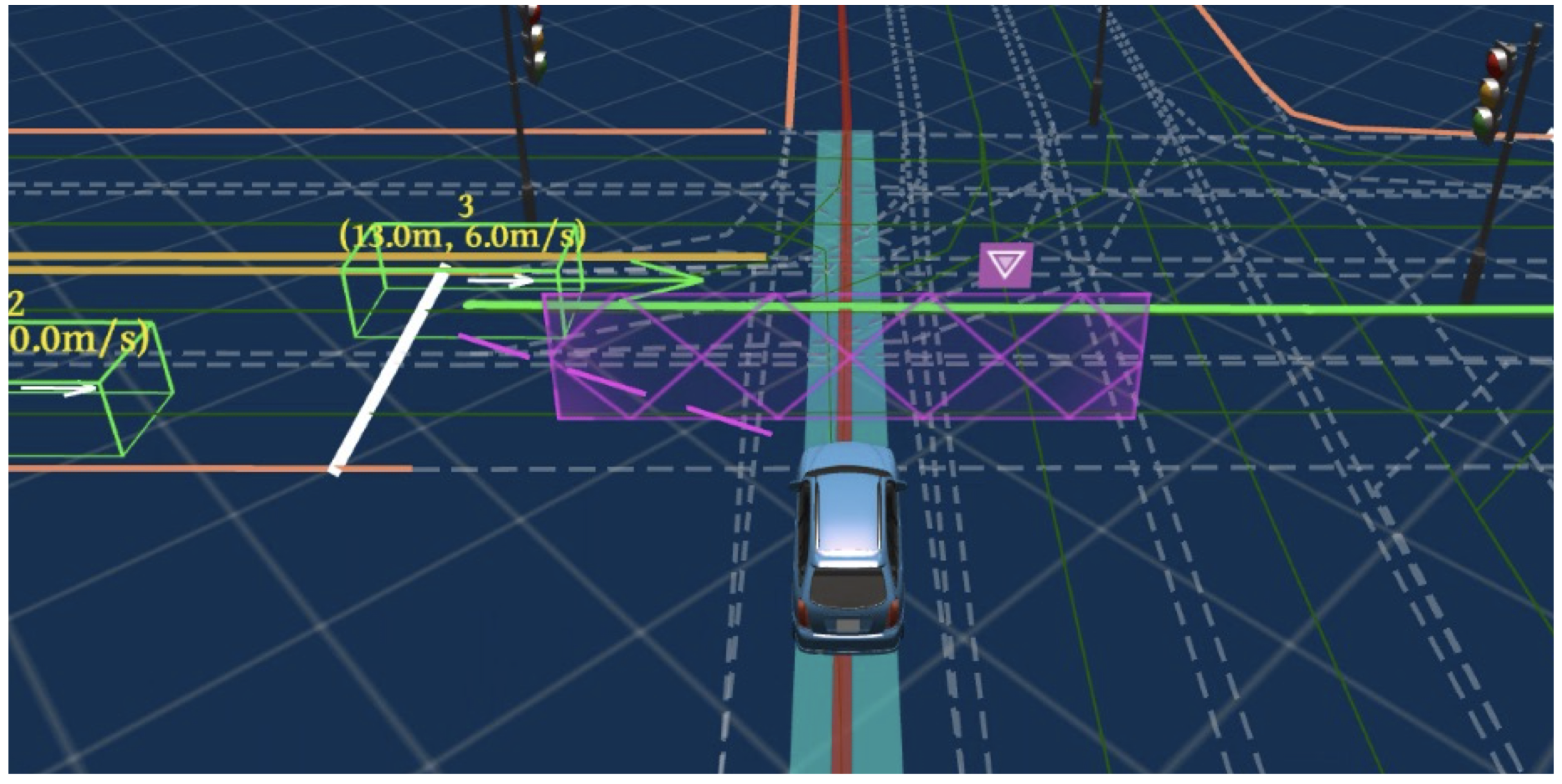}
        \caption{\var{yd}=5. Collision.}
        \label{fig:case-ydist-5}
    \end{subfigure}
    \begin{subfigure}[t]{0.31\linewidth}
        \centering
        \includegraphics[width=\linewidth]{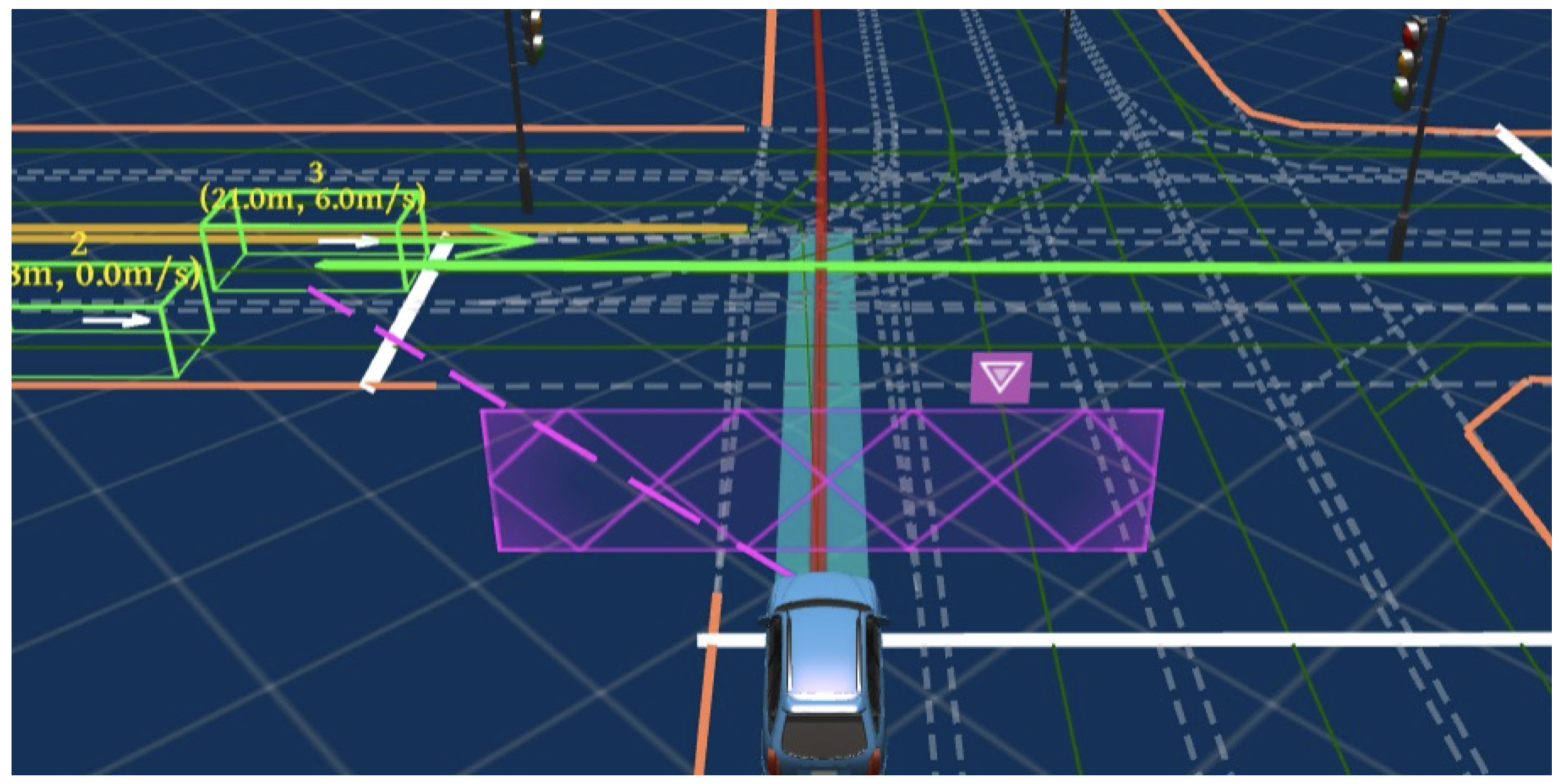}
        \caption{\var{yd}=10. Collision.}
        \label{fig:case-ydist-10}
    \end{subfigure}
    \begin{subfigure}[t]{0.31\linewidth}
        \centering
        \includegraphics[width=\linewidth]{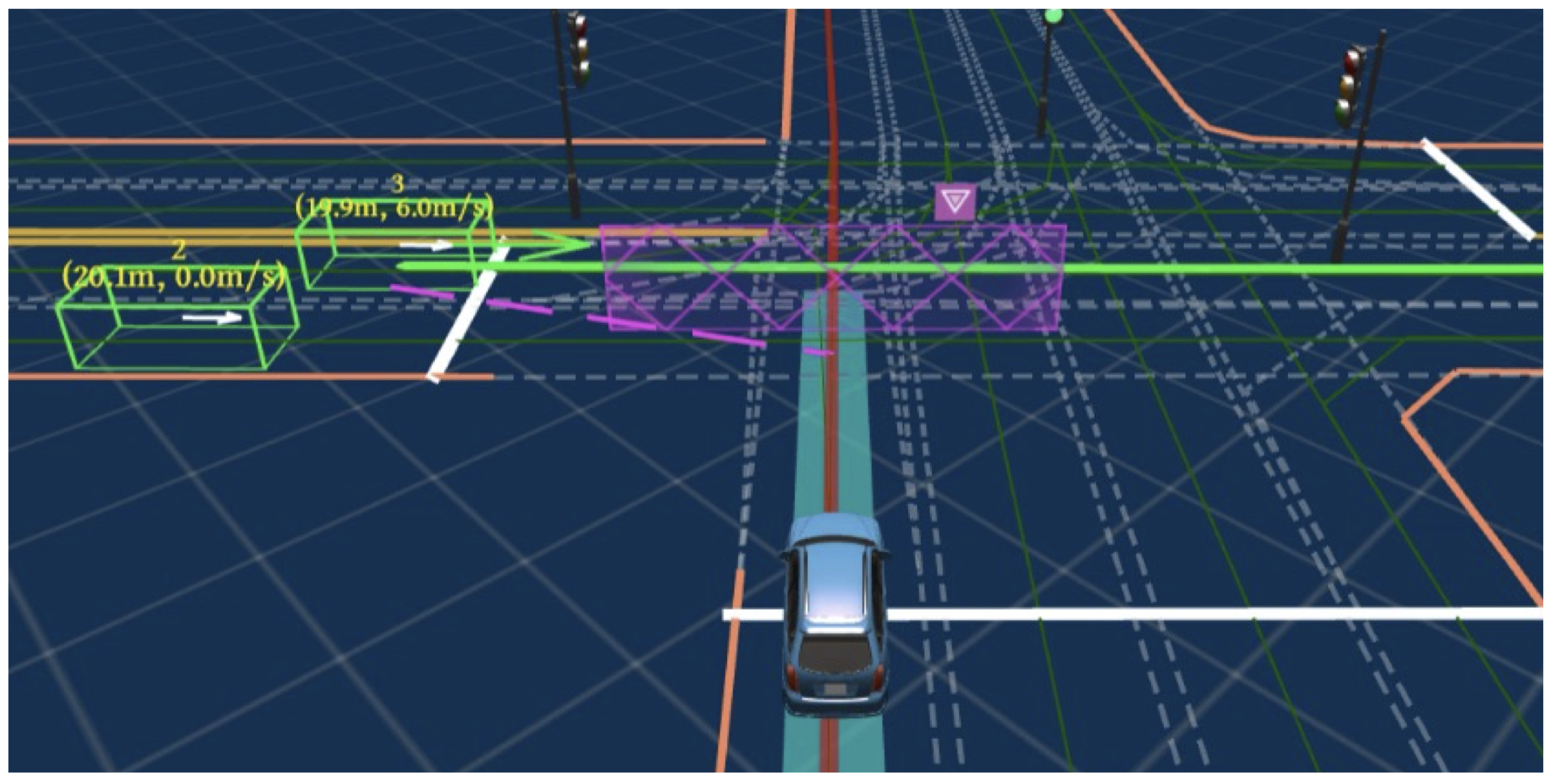}
        \caption{\textls[-10]{\var{yd}=2.No~collision.}}
        \label{fig:case-ydist-2}
    \end{subfigure}
    \caption{Replay Accident A4 with different values for \var{yield\_distance} (\var{yd} for short)}
    \label{fig:case-ydist}
\end{figure*}

\begin{figure*}[h]
    \centering
    \footnotesize
    \begin{minipage}{0.93\textwidth}
\begin{lstlisting}[lbc = {\btLstHL{4}}]
const double yield_distance = 5.0
bool SpeedDecider::CreateYieldDecision(Obstacle& obstacle, 
    ObjectDecisionType* yield_decision) {
      yield_decision->set_distance_s(-yield_distance); // set YIELD decision
      ...  }
\end{lstlisting}
\end{minipage}
\caption{Simplified code logic for Accident A4.}\label{code:case-ydist}
\end{figure*}

\noindent \textbf{Case Study I: Yield Distance (Accident A4).} In A4 (as shown in Figure~\ref{fig:case-imgs}c), the subject ADS (in blue) is driving ahead and intends to pass a crossing. At the same time, another vehicle (in red) is attempting a left turn into the same lane as the ADS. Unexpectedly, the ADS fails to yield to the red vehicle and collides with it.

To investigate the root cause of this accident, \toolname first mutates the surrounding entities (including the red vehicle) to suppress the accident. \toolname finds that the collision is prevented when the speed of the red vehicle is reduced from 6 m/s to 4 m/s. 
Next, \toolname conducts the differential analysis on the executions with and without the accident and pinpoint to the planning module.
\toolname then mutates configurations within the planning module and finds that the collision can be suppressed by changing the value of \var{yield\_distance}. Figure~\ref{fig:case-ydist} illustrates different execution results with different \var{yield\_distance} values. In the original execution with the accident, \var{yield\_distance} is set to 5 (Figure~\ref{fig:case-ydist-5}), and the collision still occurs when the value is increased to 10 (Figure~\ref{fig:case-ydist-10}). However, the collision is avoided when the \var{yield\_distance} is decreased to 2 (Figure~\ref{fig:case-ydist-2}). It indicates that this vulnerability is related to the values of \var{yield\_distance}. 

To further investigate the root case, we locate the usage of \var{yield\_distance} in source code. Figure~\ref{code:case-ydist} shows the code snippet of \var{CreateYieldDecision} API, which is used to make the yield decision. In Line 4, parameter \var{yield\_distance} is used to compute the location point at which the AD vehicle plans to stop for yielding ahead vehicles (e.g., the red one). When the value is too large, such as in Figure~\ref{fig:case-ydist-5} and~\ref{fig:case-ydist-10}, the stop location is too close for the subject AD vehicle to make a full stop. This causes the vehicle to make a plan of not yielding and thus triggers the collision. This vulnerability can be fixed by setting the parameter \var{yield\_distance} to a smaller value, such as 2 used in Figure~\ref{fig:case-ydist-2}.

%% file: figtex/case-study2.tex
\smallskip

\begin{figure*}[h]
    \centering
    \begin{subfigure}[t]{0.31\linewidth}
        \centering
        \includegraphics[width=\linewidth]{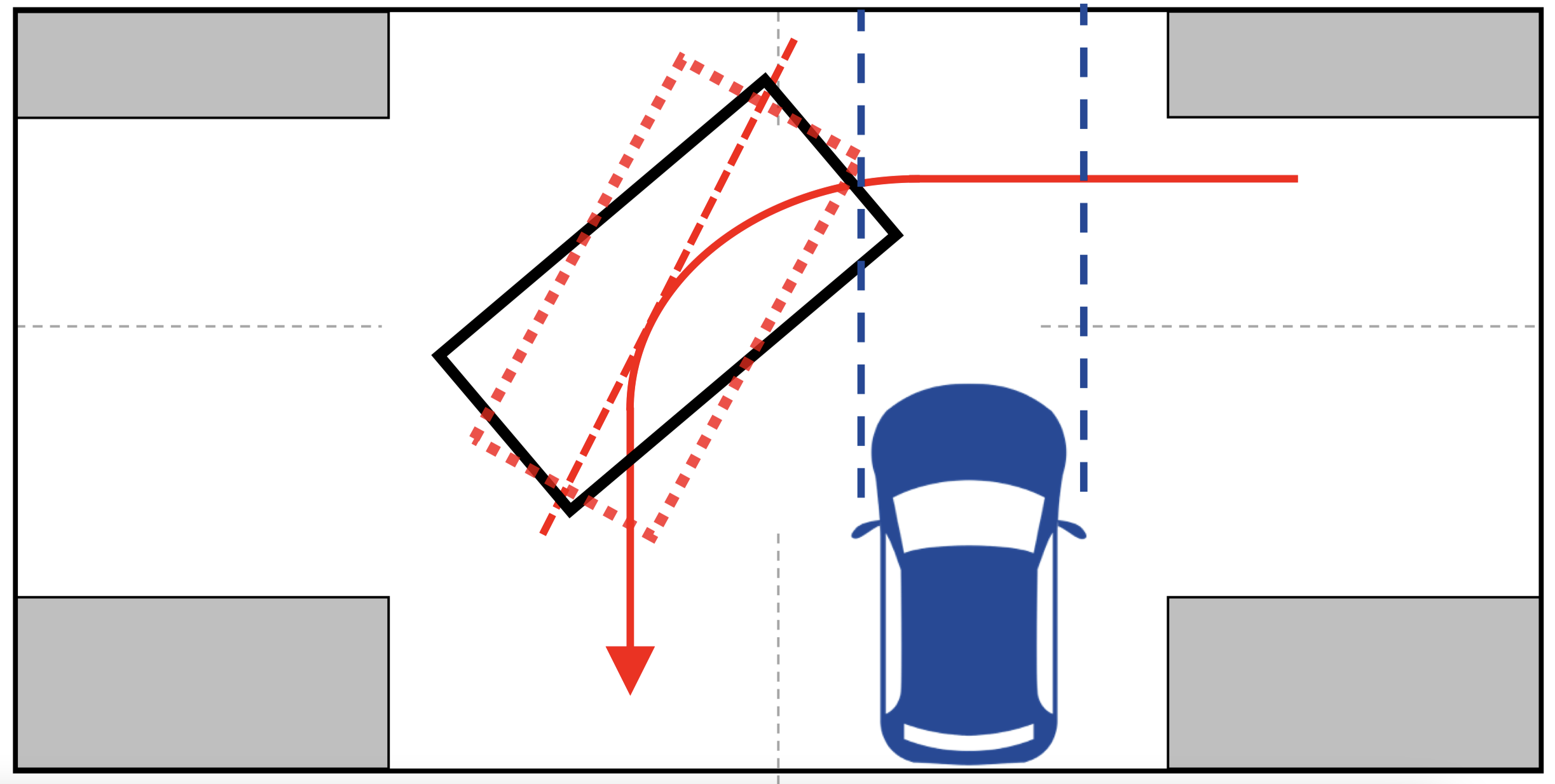}
        \caption{Illustration. }
        \label{fig:case-truck-illus}
    \end{subfigure}
    \begin{subfigure}[t]{0.31\linewidth}
        \centering
        \includegraphics[width=\linewidth]{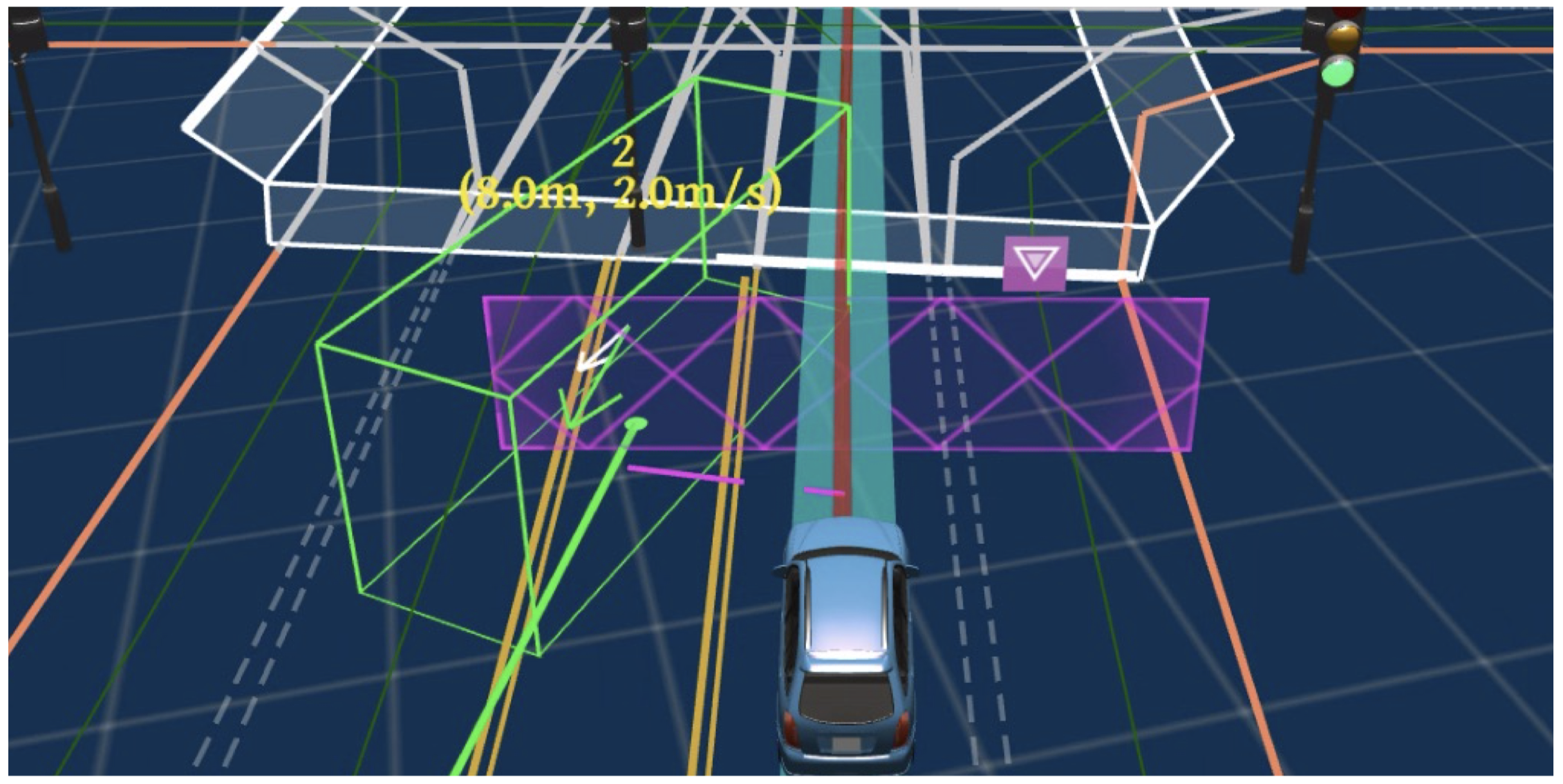}
        \caption{\var{Path.theta}. v = 10 km/h. Collision.  }
        \label{fig:case-truck-before}
    \end{subfigure}
    \begin{subfigure}[t]{0.31\linewidth}
        \centering
        \includegraphics[width=\linewidth]{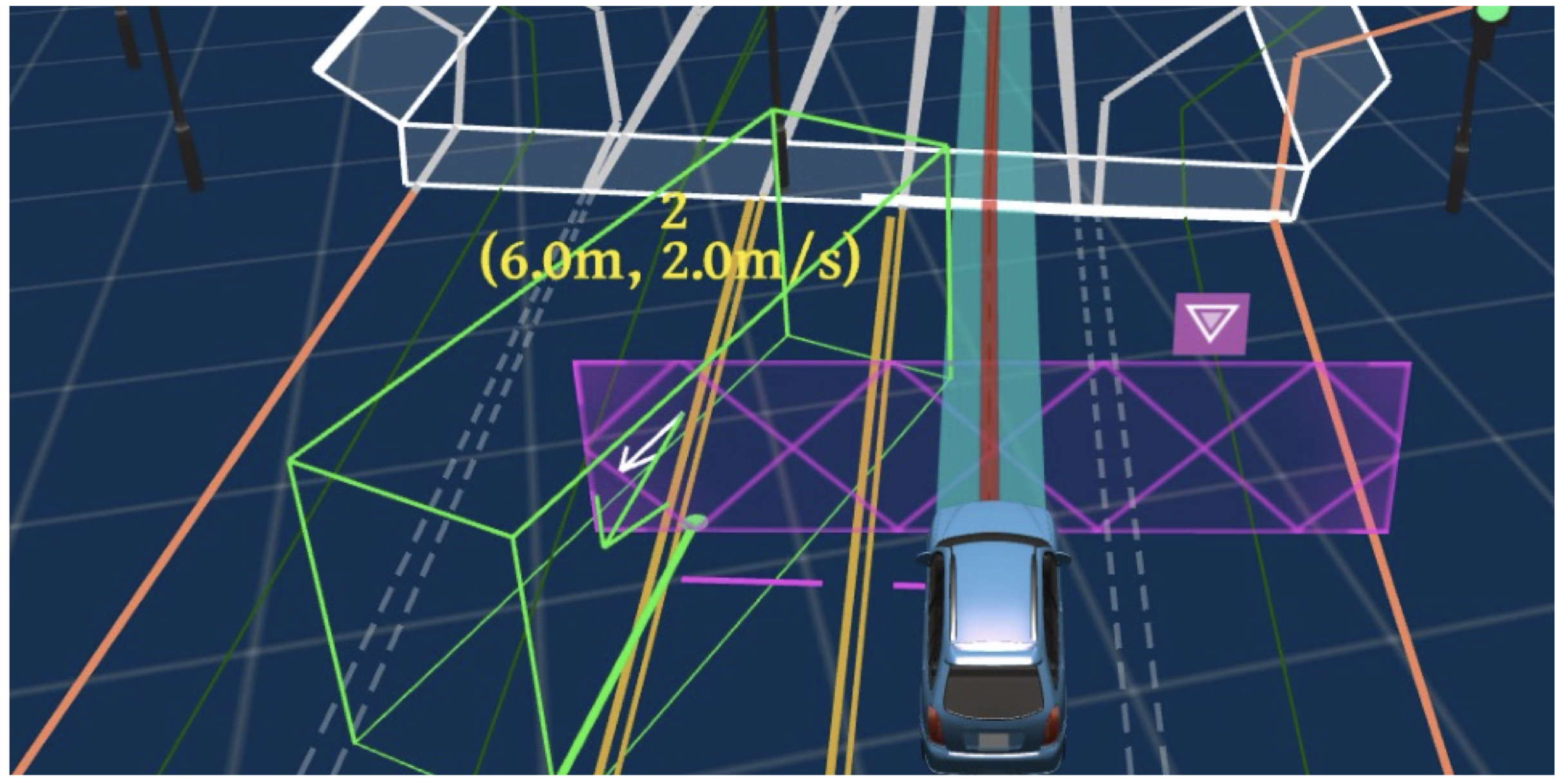}
        \caption{\var{Obs.theta}. v= 0 km/h. No collision.}
        \label{fig:case-truck-after}
    \end{subfigure}
    \caption{Replay Accident A6. In Figure~\ref{fig:case-truck-illus}, black box is the truck. Red dotted box is what AD vehicle thinks.}
    \label{fig:case-truck}
\end{figure*}
\begin{figure*}[h]
\centering
\begin{minipage}{0.93\textwidth}	
\scriptsize
\begin{lstlisting}[lbc = {\btLstHL{4, 10}}]
const double kADCSafetyLBuffer = 0.1
Box2d Obstacle::GetBoundingBox(TrajectoryPoint& point) const {
  return Box2d(point.path_point().x(), point.path_point().y(),
                    point.path_point().theta(), ...);  }
bool ComputeObstacleSTBoundary(Obstacle& obs, ...) {
  auto& obs_traj = obstacle.Trajectory();
  for (auto& obs_tjpt : obs_traj.traj_point()) {
    Box2d& obs_box = obs.GetBoundingBox(obs_tjpt);
    // adc_path_points is AD Car's planning path
    if (GetOverlappingS(adc_path_points, obs_box, kADCSafetyLBuffer)) {...} 
  } }
\end{lstlisting}
\end{minipage}
\caption{Simplified code logic for Accident  A6}\label{code:case-truck}
\end{figure*}

\noindent \textbf{Case Study II: Drifting Truck (Accident A6).}
Accident A6 (Figure~\ref{fig:case-imgs}e) depicts a more complicated accident case. As shown in the simulator screenshot, this accident occurs on a rainy day. The AD vehicle is driving ahead and passing a crossing when an external truck (in red) is turning left at the intersection. However, the ADS fails to yield the truck and collides with it.

To identify the vulnerability that causes this accident, \toolname first replays the accident in the simulator and performs the physical mutation. By mutating the property \var{Rotation.yaw} from 310 to 300, \toolname is able to find an accident-free execution. \var{Rotation.yaw} represents the angle between the ahead truck and the path that the ADS is traveling. Therefore, when the position of the truck is rotated to be outside the ADS’s path, the collision can be avoided. Given this accident-free execution, \toolname compares it with the accident execution that results in the accident and localizes the planning module as the problematic module. \toolname then performs the cyber mutation on the configuration space within the planning module and finds that the accident can be suppressed by increasing the configuration \var{kADCSafetyLBuffer} from 0.1 to 1.

To investigate the root cause of the vulnerability, we examine the usage of \var{kADCSafetyLBuffer} in the source code. As shown in Figure~\ref{code:case-truck}, \var{kADCSafetyLBuffer} is used in the API \var{\seqsplit{ComputeObstacleSTBoundary}} (Line 5), which computes the boundary of external obstacles. Specifically, it computes the obstacle’s bounding box based on its trajectory (Line 6 and Line 8) and determines whether the ADS will overlap with the obstacle’s bounding box \var{obs\_box} (Line 10). The configurable parameter \var{kADCSafetyLBuffer} is used to extend the overlap detection area by adding a buffer to its width for increased safety. A larger \var{kADCSafetyLBuffer} means the ADS detects external obstacles in a wider range. While we find that increasing \var{\seqsplit{kADCSafetyLBuffer}} can prevent the accident in this case, it may not be an optimal solution, as it could potentially trigger other vulnerabilities in different scenarios, such as making an unexpected stop or yield for an external obstacle with a safe distance. 

Note that \var{kADCSafetyLBuffer} is used for additional safety consideration. Even when its value is set to zero, the collision should not occur as long as the obstacle is outside the ADS’s path. This suggests that the vulnerability is likely caused by using an incorrect obstacle bounding box, i.e., \var{obs\_box}, another variable considered in overlap checking (Line 10). The results of physical mutation also confirm this hypothesis, where \toolname finds that adjusting the truck’s direction can avoid the accident. \var{obs\_box} is obtained by calling \var{GetBoundingBox} (Line 8),
defined at Line 2.
Upon investigation, we find that a bounding box contains information such as the position (x and y in Line 3) and the angle (theta in Line 4). However, the code uses the tangent to the path as the angle (i.e., \var{path\_point().theta()} in Line 4), instead of the real angle of the truck. We illustrate the difference between the two values in Figure~\ref{fig:case-truck-illus}. The red curve shows the trajectory of the truck, and the red box in dotted line shows the computed bounding box by the ADS, which has no overlap with the path of the ADS. However, due to the rainy weather, the large truck drifts on wet ground during the turn, making its real angle different from the tangent to the path. Thus, the real bounding box (the black box) is different from the computed one, which actually overlaps with the path of the ADS. It indicates that the vulnerability can be fixed by replacing \var{path\_point().theta()} with \var{obs.theta()}. When using \var{path\_point().theta()} in the original execution (Figure~\ref{fig:case-truck-before}), the ADS does not yield and collides with the truck at a velocity of 10 km/h. However, when using \var{obs.theta()} (Figure~\ref{fig:case-truck-after}), the ADS detects the truck successfully and stops completely, waiting for the truck to bypass.

%% file: tables/case_stats.tex
\begin{table*}[h]
    \centering 
    \footnotesize
    \tabcolsep=3pt

    \caption{Accident Statistics. We classify accidents from three perspectives: {\ding{182}} exploitability, {\ding{183}} consequences, and {\ding{184}} driving scenarios. We specifically exclude accidents that are not exploitable or only result in minor consequences (e.g., inefficiency such as taking longer routes), categorizing the remaining cases.
    }
    \label{tab:case-stats}
    \begin{tabular}{lccccccccc}
         \toprule
         \multirow{2.5}{*}{\bf Accident Sources} & 
         \multirow{2.5}{*}{\bf $\#$Accident} & 
         \multirow{2.5}{*}{\bf {\ding{182}} Exploitable} &
         \multicolumn{3}{c}{{\bf {\ding{183}} Consequence}} & 
         \multicolumn{4}{c}{{\bf {\ding{184}} Driving Scenario}} 
         \\
         \cmidrule(lr){4-6} \cmidrule(lr){7-10}
         & & &{\bf Collision} &{\bf Emergency Braking} & {\bf Inefficiency}& \textbf{Intersection} & \textbf{Merging} & \textbf{Following} & \textbf{Turning} \\ %
         \midrule
         CROUTINE~\cite{tang2021icra} & 6 & 3 (50.0\%) & 1 (16.7\%) & - & 5 (83.3\%) & 
         - & - & 3 (50.0\%) & 3 (50.0\%) \\ %
         
         CAT~\cite{tang2021iv} & 5 & 5 (100\%) & 2 (40.0\%) & 1 (20.0\%) & 2 (40.0\%) & 
         4 (80.0\%) & - & 1 (20.0\%) & - \\ %
         
         ATLAS~\cite{tang2021ase} & 10 & 7 (70.0\%) & 4 (40.0\%) & 2 (20.0\%) & 4 (40.0\%) & 
         9 (90.0\%) & - & 1 (10.0\%) & -\\ %
         
         DriveFuzz~\cite{drivefuzz} & 31 & 21 (67.7\%) & 18 (58.1\%) & 4 (12.9\%) & 9 (29.0\%) & 2 (6.5\%) & 4 (12.9\%) & 16 (51.6\%) & 9 (29.0\%)\\ %
         
         DoppelTest~\cite{huai2023doppeltest} & 123 & 112 (91.1\%) & 98 (79.7\%) & 23 (18.7\%) & 2 (1.6\%) & 40 (32.5\%) & 81 (65.9\%) & 2 (1.6\%) & -\\ %
         
         PlanFuzz~\cite{planfuzz} & 9 & 9 (100\%) & - & 6 (66.7\%) & 3 (33.3\%) & 
         3 (33.3\%) & - & 6 (66.7\%) & -\\ %
         \midrule
         \textbf{Summary} & 184 & 157 (85.3\%) & 123 (66.8\%) & 36 (19.6\%) & 25 (13.6\%) & 58 (31.5\%) & 85 (46.2\%) & 29 (15.8\%) & 12 (6.5\%) \\
         \bottomrule
    \end{tabular}
\end{table*}